\documentclass[]{aa} 

\usepackage{natbib,twoopt}
\usepackage[varg]{txfonts}
\usepackage{hyperref} 
\usepackage{multicol}
\usepackage{multirow}
\usepackage{textcomp} 
\usepackage{pifont}   
\usepackage{booktabs}
\usepackage{amssymb}
\usepackage{amsmath}
\graphicspath{{images/}{../images/}{../../images/}}
\usepackage{graphicx}
\usepackage{geometry}
\usepackage{physics}
\usepackage[math]{cellspace}
\usepackage{tabu}
\usepackage{gensymb}
\usepackage{tikz}

\usepackage{bm}

\hypersetup{
    colorlinks=true,
    urlcolor=blue,
    linkcolor=blue,
    citecolor=blue,
}
\defcitealias{1978Ap&SS..56..285L}{LD78}
\defcitealias{2020arXiv201013905L}{LB20}
\bibpunct{(}{)}{;}{a}{}{,}             
\title{Detection of nonlinear resonances among gravity modes of slowly pulsating B stars: results from five iterative prewhitening strategies}
\subtitle{}

\author{
J.~Van~Beeck \inst{\ref{kul} \href{https://orcid.org/0000-0002-5082-3887}{\includegraphics[width=3mm]{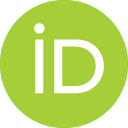}}} 
\and D.~M.~Bowman \inst{\ref{kul} \href{https://orcid.org/0000-0001-7402-3852}{\includegraphics[width=3mm]{Images/labels/orcid_128x128.png}}} 
\and M.~G.~Pedersen \inst{\ref{Kavli} \href{https://orcid.org/0000-0002-7950-0061}{\includegraphics[width=3mm]{Images/labels/orcid_128x128.png}}} 
\and T.~Van~Reeth \inst{\ref{kul} \href{https://orcid.org/0000-0003-2771-1745}{\includegraphics[width=3mm]{Images/labels/orcid_128x128.png}}}
\and T.~Van~Hoolst \inst{\ref{kul},\ref{rob}  \href{https://orcid.org/0000-0002-9820-8584}{\includegraphics[width=3mm]{Images/labels/orcid_128x128.png}}}
\and C.~Aerts \inst{\ref{kul},\ref{nijm},\ref{MPIA}  \href{https://orcid.org/0000-0003-1822-7126}{\includegraphics[width=3mm]{Images/labels/orcid_128x128.png}}}} 
\institute{Institute of Astronomy, KU Leuven, Celestijnenlaan 200D, 3001 Leuven, Belgium\\ e-mail: \texttt{jordan.vanbeeck@kuleuven.be} \label{kul} 
\and Kavli Institute for Theoretical Physics, University of California, Santa Barbara, CA 93106, USA \label{Kavli}
\and Reference Systems and Planetology, Royal Observatory of Belgium, Brussels, Belgium \label{rob}
\and Dept. of Astrophysics, IMAPP, Radboud University Nijmegen, 6500 GL, Nijmegen, The Netherlands \label{nijm} 
\and Max Planck Institute for Astronomy, Koenigstuhl 17, 69117 Heidelberg, Germany \label{MPIA}} 

\date{Received June 17, 2021 / Accepted August 5, 2021}

\titlerunning{Nonlinear g-mode resonances in SPB stars}
\authorrunning{J. Van Beeck et al.}

\abstract {Slowly pulsating B (SPB) stars are main-sequence multi-periodic oscillators that display non-radial gravity modes. For a fraction of these pulsators, 4-year photometric light curves obtained with the {\em Kepler} space telescope reveal period spacing patterns from which their internal rotation and mixing can be inferred. In this inference, any direct resonant mode coupling has usually been ignored so far.} {We re-analysed the light curves of a sample of 38 known {\em Kepler} SPB stars. For 26 of those, the internal structure, including rotation and mixing, was recently inferred from their dipole prograde oscillation modes. Our aim is to detect direct nonlinear resonant mode coupling among the largest-amplitude gravity modes.} {We extract up to 200 periodic signals per star with five different iterative prewhitening strategies based on linear and nonlinear regression applied to the light curves. We then identify candidate coupled gravity modes by verifying whether they fulfil resonant phase relations.} {For 32 of 38 SPB stars we find at least 1 candidate resonance that is detected in both the linear and the best nonlinear regression model fit to the light curve and involves at least one of the two largest-amplitude modes.} {The majority of the {\em Kepler} SPB stars reveal direct nonlinear resonances based on the largest-amplitude modes. These stars are thus prime targets for nonlinear asteroseismic modelling of intermediate-mass dwarfs to assess the importance of mode couplings in probing their internal physics.}
\keywords{asteroseismology -- stars: oscillations (including pulsations) -- stars: early-type -- stars: variables: general -- stars: rotation -- methods: data analysis}

\begin{document}
\maketitle
\section{Introduction}\label{sec:1}
Three decades ago \citet{1991A&A...246..453W} first coined the group of ``slowly pulsating B'' (SPB) stars to describe mid-B variable stars that display low-frequency oscillations with physical properties similar to 53 Persei variables \cite[see e.g.,][and references therein]{1985A&A...152....6W}. 
The SPB stars are mid-to-late B main-sequence (MS) stars with masses ranging from $\sim 3$~to~$\sim 9$~M$_\sun$ \citep{2002A&A...393..965D, 2010aste.book.....A, 2021NatAs.tmp...80P}  displaying all levels of rotation, from very slow up to the critical rotation rate \citep{2021NatAs.tmp...80P}.
A distinct feature is their multi-periodic photometric and spectroscopic variability, with periods typically ranging from $\sim 0.5$ to $\sim 5.0$~d and amplitudes typically less than $\sim 10$~mmag \cite[e.g.,][]{2010aste.book.....A, 2017A&A...598A..74P, 2021NatAs.tmp...80P}. Their variability is attributed to low-degree, high-radial order gravity ($g$) modes, which are excited by the heat engine (i.e.\ $\kappa$) mechanism driven by an opacity enhancement from the iron-group elements, also called the Z Bump, at a temperature of $\sim 200~000$ K \citep{1993MNRAS.262..213G, 1993MNRAS.265..588D, 1999AcA....49..119P}.

The oscillations in SPB stars are fairly well-characterised from ground-based photometric and spectroscopic observations \citep[e.g.,][]{1999A&A...343..872A,2000A&A...355.1015D,2001A&A...379..905M,2002A&A...393..965D,2005A&A...432.1013D,2007A&A...463..243D}, but their asteroseismic potential only became clear when {CoRoT} \citep{2009A&A...506..411A} and {\em Kepler} \citep{2010Sci...327..977B} space photometry became available \cite[see e.g.,][for recent examples of light curves]{2010Natur.464..259D,2011MNRAS.413.2403B, 2015MNRAS.451.1445B, 2012A&A...542A..55P, 2014A&A...570A...8P, 2015ApJ...803L..25P, 2017A&A...598A..74P, 2018MNRAS.478.2243S, 2021NatAs.tmp...80P, 2021MNRAS.503.5894S}.
Because of their wealth in number of excited
and identified oscillation modes, they have been the target of various asteroseismic modelling efforts, which focused on inferring the internal mixing and rotation \cite[e.g.,][]{2015A&A...580A..27M, 2016ApJ...823..130M, 2015ApJ...810...16T, 2017A&A...598A..74P,2021NatAs.tmp...80P, 2021A&A...650A.175M} and on mode excitation \citep[e.g.,][]{2018MNRAS.478.2243S, 2021MNRAS.503.5894S}. There have also been efforts to characterise the magnetic field of a small number of SPB stars (e.g. HD~43317) by analysing the frequency shifts of its $g$-modes \citep{Buysschaert2018c, 2019A&A...627A..64P, 2020A&A...636A.100P}. The influence of rotation on the $g$~modes in SPB stars is more important than the magnetic shifts and imply that 
almost all modes are in the gravito-inertial regime \citep{2017ApJ...847L...7A,2019ARA&A..57...35A}. The effects of rotation, therefore, cannot be incorporated from a perturbative treatment of the Coriolis acceleration, which led many authors to adopt the so-called traditional approximation of rotation \cite[TAR; see e.g.,][]{1968RSPTA.262..511L, 1997ApJ...491..839L, 2003MNRAS.340.1020T,2009A&A...506..111D,2013LNP...865...23M} for
pulsation computations \cite[e.g.,][]{2018MNRAS.478.2243S, 2021NatAs.tmp...80P, 2021MNRAS.503.5894S, 2021A&A...650A.175M}. 

Rapid rotation also deforms a star, causing it to become oblate. The rotation rates of SPB stars cover the cases from slight to major deformation. However, the detected and identified modes have their dominant mode energy deep inside the star, close to the convective core, where spherical symmetry still applies well. This  
justifies a treatment of g~modes in the TAR \citep{2019A&A...631A..26M,2021A&A...648A..97H,2021arXiv210409302D}. An important consequence of the rapid rotation of some SPB stars is the occurrence of mode coupling between overstable convective modes based on frozen-in core convection and envelope $g$ modes excited by the $\kappa$ mechanism \citep{2021MNRAS.505.1495L}.

SPB stars have so far been modelled from linear pulsation theory. Period spacing patterns constructed from the frequencies of identified $g$ modes of consecutive radial order and the same angular degree and azimuthal order, are the most important tool used in linear asteroseismic modelling. Such patterns can display oscillatory features attributable to chemical gradients inside the star \cite[e.g.,][]{2008MNRAS.386.1487M}. Moreover, they display a slope due to the rotation rate near the convective core \cite[e.g.,][]{2013MNRAS.429.2500B, 2017A&A...598A..74P}. Some SPB stars, however, can show structure in their g-mode period spacing patterns that is hard to explain without incorporating additional physical effects  in the oscillation model, such as rotational deformation \citep[e.g.,][]{2021arXiv210409302D} or (internal) magnetism \cite[see e.g.,][]{2020A&A...638A.149V}, or resonances of envelope $g$ modes with inertial modes in the core \citep[e.g.,][]{2020A&A...640A..49O, 2021MNRAS.502.5856S}.

The input to asteroseismic modelling based on linear oscillation theory are the independent frequencies extracted from an observed time series (e.g., a light curve), where frequency extraction is usually done using an iterative prewhitening procedure \cite[e.g.,][]{2009A&A...506..111D}. However, non-sinusoidal light curves give rise to harmonics and combination frequencies (dependent frequencies) in the amplitude spectrum \citep{2012AN....333.1053P,2015MNRAS.450.3015K}. These combination frequencies are therefore also included in the extracted frequencies. The phases and amplitudes of the independent frequencies, as well as the frequencies, amplitudes, and phases of the combination frequencies are subsequently neglected during linear asteroseismic modelling. 
On the other hand, frequency perturbations within g-mode period spacing patterns are expected if nonlinear mode coupling is accounted for \cite[e.g.,][]{1984ApJ...279..394B, 1996A&A...308...66V}. Hence, this may be an alternative explanation for the deviations in the period spacing patterns of SPB stars. Such frequency shifts may be small if the $g$~modes are only moderately nonlinear, as is the case for most `well-observed' nonradial pulsators such as white dwarfs or $\delta$ Sc stars \citep{1997A&A...321..159B}, yet this remains to be verified for SPB stars.

As detailed in \citet{1997A&A...321..159B}, resonances among (excited) non-radial oscillation modes also play an important role in mode selection and interaction, and can lead to additional constraints on mode identification. Resonant interactions among modes can further hamper pattern identification due to so-called frequency or phase locking. To complicate pattern identification the exact resonance relation, $\sum_i n_i\, \omega_i = 0$, need not be satisfied for linear mode frequencies. It suffices that linear mode frequencies are in near resonance, such that the exact resonance relation between angular frequencies $\omega_j$ is satisfied aside from a small angular frequency difference $\delta\omega$. Nonlinear frequency shifts can then lead to a nonlinear frequency locking or phase locking, where the nonlinear locked frequencies satisfy the resonance condition exactly, which can distort the period spacing patterns relied upon in linear asteroseismology. 

The theoretical framework of resonant mode coupling, the amplitude equation (AE) formalism, is described in numerous publications \cite[e.g.,][]{1982AcA....32..147D, 1984ApJ...279..394B, 1985AcA....35....5D, 1993Ap&SS.210....9B, 1997A&A...321..159B, 1993A&A...279..417V, 1994A&A...292..471V, 1994A&A...286..879V, 1995A&A...295..371V}. It has the inherent validity assumption that the (linear) growth rates, $\gamma$, of the modes are small compared to their angular frequencies, which is generally assumed to be satisfied for observed MS oscillators \cite[see e.g.,][]{2015A&A...579A.133B, 2016MNRAS.460.1970B}.
The AEs constitute a set of coupled first-order complex nonlinear ordinary differential equations and predict the temporal behaviour of the mode amplitudes and phases. Three main regimes of mode interaction are distinguished \cite[as explained in][]{1997A&A...321..159B}:
\begin{enumerate}
    \item If two or more modes are close to being in resonance, such that $D_g \equiv \delta\omega / \gamma$ is of order $1$ or smaller (where $\delta \omega$ is the angular frequency mismatch or angular frequency detuning), and a stable fixed point (FP) solution exists for the AEs, for which the mode amplitudes and phases are constant over time, the frequencies are locked. These locked frequencies can be substantially different from their non-resonant counterparts. 
    
    \item With an increasing angular frequency mismatch, $\delta \omega$, the FP solution becomes unstable or disappears, and a bifurcation to another solution takes place, with amplitude (and phase) modulation. The larger the value of $D_g$, the smaller the amplitude of the modulation. This is called the `intermediate' regime \citep{1998BaltA...7...21G}.
    
    \item At large $\delta \omega$ one encounters the solution where frequencies are approximately equal to their (unlocked) non-resonant values. The nonlinear nonresonant frequencies have a small frequency shift compared to their linear counterparts: this is the mildly nonlinear regime in which linear asteroseismology operates, also called the `nonresonant' regime \citep{1998BaltA...7...21G}.
\end{enumerate}
In addition, a narrow hysteresis (transitory) regime exists in between the frequency-locked and `intermediate' regime, in which amplitudes are modulated, but frequencies (or equivalently, phases) remain constant \citep{1997A&A...321..159B}.

The study of the relationships among amplitudes and phases, which provide the observables to be matched to the theoretical predictions from the AEs, is a standard approach for the analysis of light curves of RR~Lyr stars and Cepheids \cite[as pioneered by][]{1981ApJ...248..291S}. It is also a well-established technique for unravelling nonlinearities in the light curves of oscillating white dwarf stars \cite[e.g.,][]{2001MNRAS.323..248W, 2016A&A...585A..22Z}, and hot B subdwarf stars \cite[e.g.,][]{2016A&A...594A..46Z}. Yet, it has not received much attention for oscillating MS stars \citep{2012MNRAS.420.2387L, 2014MNRAS.444.1909B, 2015MNRAS.450.3015K, 2016MNRAS.460.1970B, 2021MNRAS.504.4039B}.
In this work, we provide the identification of candidate direct resonances among the extracted oscillation frequencies in the light curves of 38 SPB stars, and test the impact of different iterative prewhitening strategies on the results. In Sect.~\ref{sec:light_curves_and_analysis} we summarise the characteristics of our sample, the strategies we employ to analyse the light curves, and discuss how we search for candidate direct resonances. We discuss the light curve analysis strategy selection process in Sect.~\ref{sec:light_curve_regression_model_selection}, and the results of the candidate resonance search in Sect.~\ref{sec:candidate_resonance_search}. A summary of our conclusions and future prospects can be found in Sect.~\ref{sec:conclusions_prospects}.

\section{Light curves and their analysis}\label{sec:light_curves_and_analysis}

\subsection{The \citet{may_phd_thesis} SPB sample}

Our sample consists of 38 stars originally identified as SPB stars based on \emph{Kepler} photometry \citep{2011MNRAS.413.2403B, 2015MNRAS.451.1445B, 2012AJ....143..101M, 2013A&A...553A.127P, 2014A&A...570A...8P, 2015ApJ...803L..25P, 2017A&A...598A..74P, 2018ApJ...854..168Z, 2018MNRAS.478.2243S}. We use the light curves extracted from the 30-min cadence target pixel data by \citet{may_phd_thesis}, who used customised pixel masks (see \citealt{2017A&A...598A..74P} for further details). After light curve extraction, \citet{may_phd_thesis} manually removed outliers, and separately corrected each quarter for instrumental trends by fitting a low order polynomial to the data. \citet{may_phd_thesis} then converted the light curves to have flux units of ppm, and normalised them to have an average flux of zero.

Contrary to previous analyses of {\it Kepler\/} SPB stars by \citet{2018MNRAS.478.2243S},
\citet{may_phd_thesis}, \citet{2021MNRAS.503.5894S} and \citet{2021NatAs.tmp...80P}, we use five different iterative prewhitening strategies, each of which produce a final regression model fit for the light curve. Our approach is motivated by the fact that iterative prewhitening techniques vary in the literature, while it is not assessed how these various approaches influence frequency extraction results. Our comparative capacity assessment of the different iterative prewhitening procedures is an important aspect of the hunt for nonlinear resonances in SPB stars.

\subsection{SPB light curve analysis}\label{subsec:SPB_light_curve_analysis}

\begin{table}
\centering
\caption{Five iterative prewhitening strategies employed in this work.}
\label{tab:strategy_table}

\begin{tabular}{crrr}
\toprule
Strategy & Hinting & Stop criterion & Final optimisation\,\tablefootmark{a} \\
\midrule
$1$ & S/N & S/N $< 4.0$ & nonlinear\\
$2$ & A& $p_{A}$\,\tablefootmark{b} $> 0.05$ & nonlinear\\
$3$ & A& $p_{\rm LRT}$\,\tablefootmark{c} $> 0.05$& nonlinear\\
$4$ & S/N& $p_{\rm LRT}$\,\tablefootmark{c} $> 0.05$& nonlinear\\
$5$\,\tablefootmark{d} & A & $p_{\rm LRT}$\,\tablefootmark{c} $> 0.05$& linear\\
\bottomrule
\end{tabular}
\tablefoot{
Parameter hinting is either performed by selecting the highest-amplitude peak (A) in the residual LS periodogram, or the peak with the highest amplitude S/N value using a 1 d$^{-1}$ window (S/N). The selected peak is not closer than 2.5\,/\,$T$ to earlier extracted frequencies, where $T$ is the total time span of the data \citep{1978Ap&SS..56..285L}. Linear interpolation between the over-sampled frequency grid points of the Lomb-Scargle periodogram \citep{1976Ap&SS..39..447L, 1982ApJ...263..835S} is employed to determine the S/N value associated with the optimised frequency.\\
\tablefoottext{a}{Type of least-squares optimisation of the entire model at the end of each iterative prewhitening step.}
\tablefoottext{b}{The $p$-value for a Z-test based on extracted amplitude.}
\tablefoottext{c}{The $p$-value for a likelihood ratio test (LRT) based on Bayesian information criterion (BIC) values for the nested light curve regression models.}
\tablefoottext{d}{The frequency extracted during the iterative prewhitening step is optimised by selecting the largest-amplitude signal in an over-sampled LS periodogram in a 1 d$^{-1}$ region around the frequency hint.}
}
\end{table}

We employ five prewhitening strategies to derive frequencies, amplitudes and phases from the light curves, which emulate various approaches taken in the literature. The first timestamp of each light curve is subtracted from all timestamps of that light curve to provide a consistent zero point in time for phase calculation.
Strategy 1 uses a nonlinear least-squares fit of the entire multi-parameter model in the time domain at each prewhitening step \citep{2016MNRAS.460.1970B}. It requires each signal to have an amplitude S/N $\geq$ 4 \citep{1993A&A...271..482B} computed in a window of 1 d$^{-1}$ around the target frequency, and fits a model comprising a sum of sinusoids, $F(t_i)$, to the light curve:
\begin{equation} \label{eq:sum_sines}
    F(t_i) = \beta_0 + \sum_{j=1}^{n_f} A_j \sin\left[2\pi \nu_j t_i + \phi_j\right]\,,
\end{equation}
for each $t_i$, $i=1, \ldots, n_t$, with $n_t$ the total number of timestamps of the light curve. In this expression, $n_f$ is the number of fitted frequencies, $\beta_0$ is the y-intercept, $A_j$ denotes the amplitude, $\nu_j \left(\equiv \omega_j / 2 \pi\right)$ is the temporal frequency, and $\phi_j$ is the phase of fitted sinusoid $j$. 
The other four prewhitening strategies differ from this first one in the way they determine the initial parameters for the optimisation (i.e.\ parameter hinting; see the notes of Table~\ref{tab:strategy_table}), stop criterion, and whether a linear or nonlinear (multi-parameter) least-squares optimisation is performed at the end of each iterative prewhitening step. These differences are listed in Table \ref{tab:strategy_table}. 

As is commonly done, we calculate the uncertainties for the extracted parameters, under the assumption that they are uncorrelated, using
\begin{equation}\label{eq:correlation_correction}
    \hat{\sigma}_{\nu} = D\,\dfrac{\sqrt{6}\,\sigma_{ n_t}}{\pi\,\sqrt{ n_t}\,\hat{ A}\,{T}}, \hspace{0.3cm} \hat{\sigma}_{ A} = { D}\,\sqrt{\dfrac{2}{ n_t}}\sigma_{ n_t}, \hspace{0.3cm} \hat{\sigma}_{\phi} = { D}\,\sqrt{\dfrac{2}{ n_t}}\dfrac{\sigma_{ n_t}}{\hat{ A}}\,,
\end{equation}
where $\sigma_{n_t}$ is the standard deviation of the residual signal (after the regression model has been subtracted from the light curve), $\hat{A}$ is the optimised amplitude, $T$ is the total time span of the data, and where $D$ is a correction factor for the correlated nature of the light curve data, which can be estimated as the square root of the average number of consecutive data points of the same sign in the residual light curve \cite[e.g.,][]{1991MNRAS.253..198S, 1999DSSN...13...28M, 2003ASPC..292..383S, 2009A&A...506..111D}. 
During each prewhitening step we compute the Lomb-Scargle (LS) periodogram \citep{1976Ap&SS..39..447L, 1982ApJ...263..835S} of the (residual) light curve, with frequency grid step equal to $0.1\mathfrak{R}_{\nu}$ (where $\mathfrak{R}_{\nu}\equiv \tfrac{1}{T}$ is the Rayleigh limit). The start and end values of this grid are equal to $1.5\,\mathfrak{R}_{\nu}$ and the Nyquist frequency $\nu_{\rm Nyq}$ ($\nu_{\rm Nyq} = 1/(2\,\Tilde{\Delta t})$, where $\Tilde{\Delta t}$ is the median cadence of the time series), respectively.
The LS periodogram, together with a specific parameter hinting rule listed in Table \ref{tab:strategy_table}, are used to provide the preliminary amplitude and frequency hint, taking into account the frequency resolution \citep{1978Ap&SS..56..285L}, hereafter referred to as the \citetalias{1978Ap&SS..56..285L} resolution. The preliminary phase hint is subsequently obtained from a linear least-squares fit to the residual light curve, where frequencies and amplitudes in Eqn.~(\ref{eq:sum_sines}) are fixed at either their frequency and amplitude hint values (the frequency being extracted), or the values they had in the previous prewhitening step (all other frequencies). 
The phase hinting optimisation employs the Trust-region reflective least-squares regression method implemented in the Python package {\sc lmfit} \citep{matt_newville_2020_3814709}, whose exact configuration is specified in Appendix~A.
For strategies 1 to 4 the optimised parameter hints are then fed to the nonlinear least-squares optimiser at the end of this prewhitening step to optimise the frequencies, amplitudes and phases of the entire model.
For strategy 5, however, these parameter hints are only used in a linear least-squares optimisation of the amplitudes and phases, where the frequencies are fixed at the values determined from an over-sampled LS periodogram in a 1 d$^{-1}$ region around the preliminary frequency hint (signal extracted during this iterative prewhitening step), and the values they had in the previous iteration step (other signals). The Levenberg-Marquardt method implemented in {\sc lmfit} \citep{matt_newville_2020_3814709} is used for this optimisation, whose exact configuration is specified in Appendix~A. 

Each iterative prewhitening process continues until 200 frequencies have been extracted in this way, or until the specified stop criterion as in Table \ref{tab:strategy_table} is reached. 
After the frequency extraction, we verify that the amplitude of each extracted frequency is significant at the 95\% level using a Z-test, whilst taking the \citetalias{1978Ap&SS..56..285L} resolution into account. If at least one insignificant extracted frequency is detected, the last extracted insignificant frequency is removed. After that frequency is removed, the entire model is re-optimised, and a new significance check is performed. This iterative process, which we refer to as the frequency filtering step, continues until no further insignificant frequencies occur in the regression model.

The regression models for each of the prewhitening strategies that have undergone frequency filtering are then compared with one another. To quantify the quality of the regression model, we rely on the residual sum of squares, weighted by the total variance in the observed light curve and by a factor including the total number of free parameters of the regression model, $n_p$, versus $n_t$. To end up with a maximum likelihood estimator, we subsequently use the 
scaled fraction of variance, $f_{\rm sv}$, which we define as
\begin{align}
    f_{\rm sv} \equiv\ 1 - \left(\dfrac{\sum_i^{n_t}\left(y_i - F(t_i)\right)^2}{\sum_i^{n_t}\left(y_i - \Bar{y}\right)^2}\right)\left\{\dfrac{{n_t} - 1}{{n_t} - n_p}\right\}\,, \label{eq:scaled_fraction_variance}
\end{align}
where $y_i$ denotes the time series signal at timestamp $t_i$, and $\Bar{y} = \sum_i^{n_t}y_i / n_t$ is the mean signal of the time series. The higher the quality of the regression model, the closer $f_{\rm sv}$ approaches unity while we punish for a higher number of free parameters.

\subsection{Candidate resonant {\em Kepler} SPB oscillations}\label{subsec:candidate_resonant_oscillations}

In our identification of direct resonant mode interactions we focus on FP solutions of the AEs, for which frequency locking persists 
\citep{1997A&A...321..159B}. Any such solutions have parent-daughter frequency combinations that approximately satisfy the (direct) resonance relation. Our aim is to extract information on the least complicated mode interactions, hence, we use strict candidate resonance identification criteria. The first criterion requires that a frequency combination of a candidate resonance satisfies the following relation:
\begin{equation}
    \exists \nu_{\rm daughter}:\hspace{0.2cm}\delta \nu \,\equiv \nu_{\rm daughter} - \sum_i n_i\, \nu_{\rm parent,\,i} \approx 0\, \text{d}^{-1}\,, \label{eq:frequency_relation}
\end{equation}
where $\nu_{\rm daughter}$ is the daughter frequency, $\nu_{\rm parent,\,i}$ is its $i$-th parent frequency, $n_i \in \mathbb{Z}$, and where $\delta \nu$ is the frequency detuning, for which $|\delta\nu |< \nu_{\rm daughter}$ holds. This follows the parent-daughter terminology introduced in \citet{2009A&A...506..111D}. To further characterise and verify the direct candidate resonance, and following \citet{1997A&A...321..159B}, \citet{2000MNRAS.313..179V} and \citet{2000MNRAS.313..185V}, we compute relative phases and amplitudes. The second candidate resonance identification criterion then consists of verifying whether the corresponding relative phases $\Phi$ of the combinations that satisfy Eqn.~(\ref{eq:frequency_relation}) also satisfy the following (resonant) relative phase relation (see e.g., Fig.~$13$ in \citealt{1995A&A...295..371V}, in which $\Gamma^{\,0}$ and $\sigma$ denote the FP relative phase and frequency detuning, respectively, as well as Fig.~$1$ in \citealt{1997A&A...321..159B}):
\begin{equation}
    \exists k \in \left[-4,\,-3,\,\ldots,\,4\right]:\hspace{0.15cm}\Phi \equiv \phi_{\rm daughter}\,- \sum_i n_i\, \phi_{\rm parent,\,i} \approx k\cdot\dfrac{\pi}{2}\,, \label{eq:relative_phase_relation}
\end{equation}
where $\phi_{\rm daughter}$ is the extracted phase of the daughter frequency, and $\phi_{\rm parent,\,i}$ is the extracted phase of its $i$-th parent frequency.
If for a given candidate combination of parent(s) and daughter frequencies Eqns~(\ref{eq:frequency_relation}) and (\ref{eq:relative_phase_relation}) are satisfied, this pair is classified as a candidate resonance. We define the combination order $o$ of this candidate resonance as: $o = \sum_i \left|n_i\right|$. Following 
\citet{2000MNRAS.313..185V} we compute the relative amplitude of the candidate resonance, $A_r$, in the following way:
\begin{equation}
    \label{eq:relative_amplitude}
    A_r = \dfrac{A_{\rm daughter}}{\displaystyle\prod_{i} \left(A_{\rm parent,\,i}\right)^{\left|n_i\right|}}\,.
\end{equation}

In practice, we limit ourselves to identifying up to second-order combinations of parent frequencies ($o = 2$). For each potential candidate direct resonance, we then verify whether Eqn.~(\ref{eq:frequency_relation}) is satisfied to within 1 Rayleigh limit $\mathfrak{R}_{\nu}$ at the 1-$\sigma$ level:
\begin{equation}
    \exists \nu_{\rm daughter}:\hspace{0.2cm}\left|\nu_{\rm daughter} - \nu\right| - \sigma_{\nu_{\rm daughter}} - \sigma_{\nu} \leq \mathfrak{R}_{\nu} \,, \label{eq:practical_check_freq}
\end{equation}
where $\nu \equiv \sum_i n_i\, \nu_{\rm parent,\,i}$ is a combination frequency composed of other extracted frequencies, with propagated uncertainty $\sigma_{\nu}$. The comparison with $\mathfrak{R}_{\nu}$ hence measures whether $\nu$ is resolved with respect to $\nu_{\rm daughter}$. We use this practical approximation for assigning candidate resonance identifiers, to avoid use of the theoretical criterion by \citet{1997A&A...321..159B},  $|\delta\omega|/\gamma \lesssim O(1)$, setting the approximate limits of parameter space for which an FP solution with frequency locking exists, as it relies on 
model-dependent nonadiabatic oscillation calculations for $\gamma$, while we want to establish a data-driven approach.  
If Eqn.~(\ref{eq:practical_check_freq}) holds, we further verify whether the relative phase $\Phi$ of the candidate combination satisfies Eqn.~(\ref{eq:relative_phase_relation}) to within 1 $\sigma$:
\begin{equation}
    \exists k \in \left[-4,\,-3,\,\ldots,\,4\right]: \hspace{0.2cm}\Phi - \sigma_\Phi \leq k \cdot \dfrac{\pi}{2} \leq \Phi + \sigma_\Phi\,. \label{eq:practical_check_phase}
\end{equation}

\section{Results: light curve regression model capacity}\label{sec:light_curve_regression_model_selection}

In this section the resulting regression models obtained for each of the five prewhitening strategies applied to all 38 stars in our sample are compared. The metrics used for this comparison are the $f_{\rm sv}$ value defined in Eqn.~(\ref{eq:scaled_fraction_variance}), and the (residual) light curves and their LS periodogram. The $f_{\rm sv}$ metric measures how well the five regression models
describe the light curve of any given star, where we recall that 
Eqn.~(\ref{eq:sum_sines}) assumes that each oscillation mode has constant amplitude, frequency, and phase. Under this assumption and using the criteria in 
Table\,\ref{tab:strategy_table}, we define 
the best regression model as the one with the highest $f_{\rm sv}$.
Any violation of the assumed constant amplitudes, phases, or frequencies results in lower values of $f_{\rm sv}$ along with the occurrence of more harmonics and combination frequencies.

\begin{figure*}
    \centering
    \includegraphics[width=18cm]{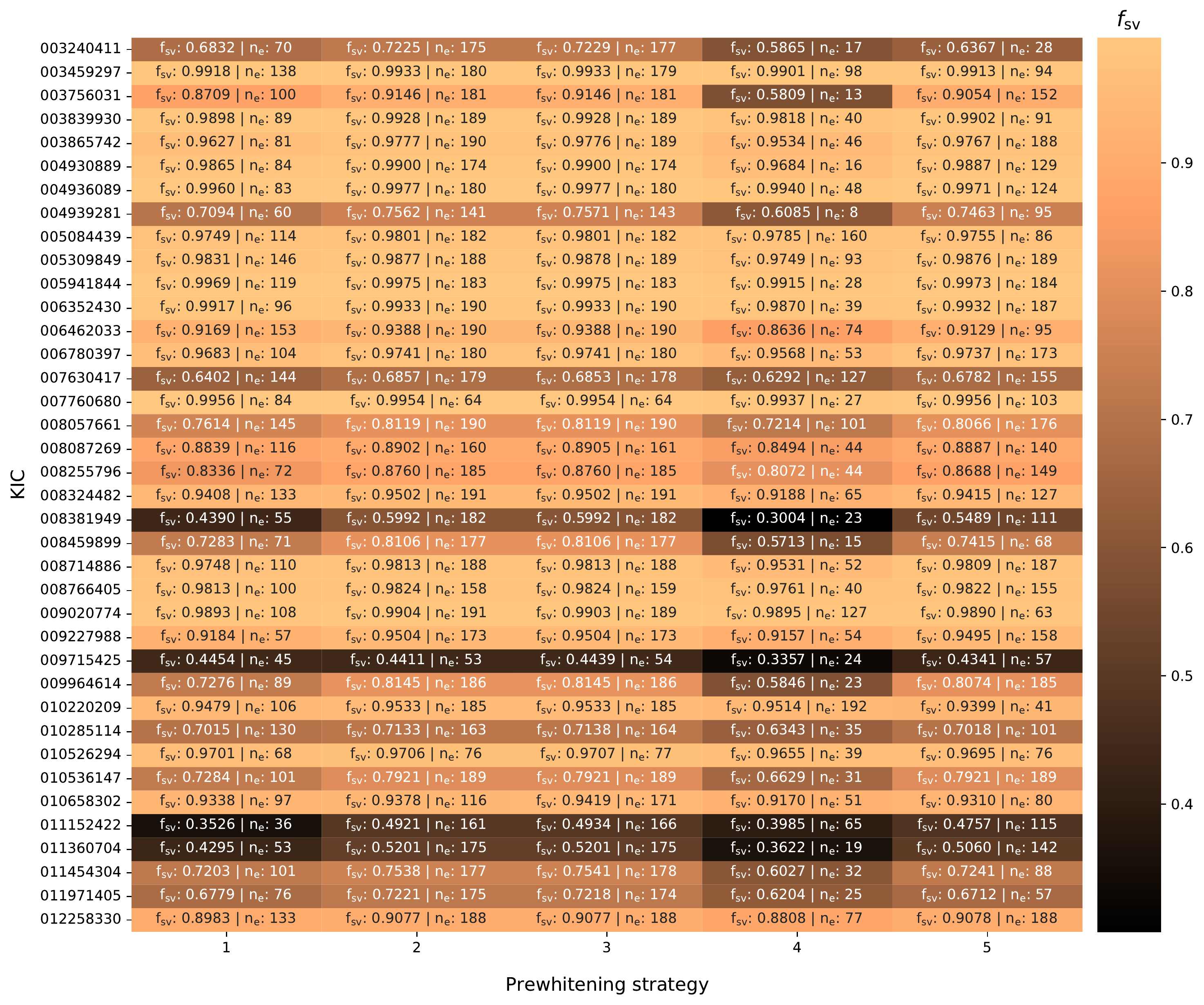}
    \caption{Comparison of the scaled fraction of variance $f_{\rm sv}$ values of the resulting light curve regression models obtained for the different prewhitening strategies applied to the stars in the sample, with colouring as a function of $f_{\rm sv}$. Each cell contains the $f_{\rm sv}$ value as computed using Eqn.~(\ref{eq:scaled_fraction_variance}), as well as the total number of frequencies extracted using this approach, $n_{\rm e}$. The number $n_{\rm e}$ and the $f_{\rm sv}$ value are computed after the frequency filtering step has removed any unresolved frequencies with respect to the \citetalias{1978Ap&SS..56..285L} criterion and any signals that have insignificant amplitudes (at $95$\% confidence), as discussed in section~\ref{subsec:SPB_light_curve_analysis}. }
    \label{fig:s_f_v_comparison_prewhitening_strategies_SPB_stars}
\end{figure*}

\begin{figure}
    \centering
    \resizebox{\hsize}{!}{\includegraphics{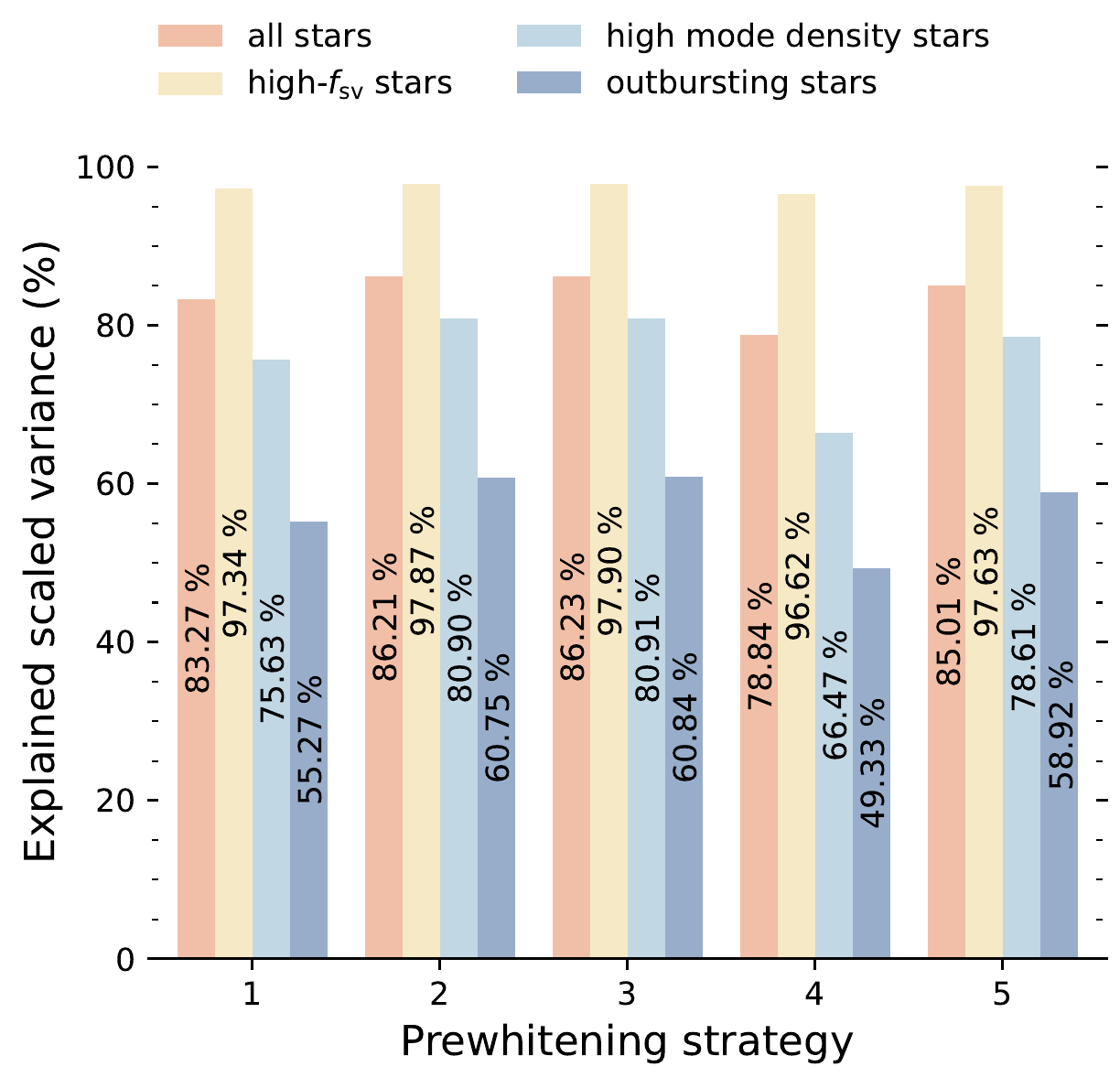}}
    \caption{Total percentage of explained scaled variance $f_{\rm sv}$ averaged over all 38 stars in the sample (`all stars') for each of the five prewhitening strategies. This averaged percentage is also displayed for each of the three pseudoclasses introduced in Section~\ref{sec:light_curve_regression_model_selection}.}
    \label{fig:explained_scaled_variance_comparison}
\end{figure}

Based on the $f_{\rm sv}$ metric, we divide the stars in our sample into two pseudoclasses: (i) stars for which the light curve regression models attain $f_{\rm sv}$ values above $0.9$ after frequency filtering for all five iterative prewhitening strategies; and (ii) stars for which at least one of these models attains a value of $f_{\rm sv} < 0.9$. For brevity, we instead refer to strategies `having' $f_{\rm sv}$ values above $0.9$ in the remainder of the text.
This classification accentuates the apparent bimodality in the distribution of the $f_{\rm sv}$ metric values that can be seen in Fig.~\ref{fig:s_f_v_comparison_prewhitening_strategies_SPB_stars}. Half of the 38 SPB stars in our sample have $f_{\rm sv} \geq 0.9$ for all five prewhitening strategies and thus belong to pseudoclass 1, hereafter referred to as the `high-$f_{\rm sv}$' stars. We discuss their light curve regression model selection process in subsection~\ref{subsec:high_s_f_v_stars}. We additionally subdivide the stars belonging the second pseudoclass (the `low-$f_{\rm sv}$' stars) into two additional pseudoclasses based on whether outburst-like features are detected in the light curves, and thus end up with three pseudoclasses that describe the whole sample. The light curve regression model selection process for these two additional pseudoclasses, the high mode density starands and the outbursting stars, are discussed in subsection~\ref{subsec:low_s_f_v_stars}, both guided by particular example stars. The percentages of explained scaled variance $f_{\rm sv}$ averaged over all stars in our sample, and averaged over the stars belonging to each of these pseudoclasses are shown in Fig.~\ref{fig:explained_scaled_variance_comparison}. According to this `sample' metric, high-$f_{\rm sv}$ stars are explained well by the regression models belonging to the five  prewhitening strategies. However, significantly differing values of $f_{\rm sv}$ can be noted for low-$f_{\rm sv}$ stars. For the sample of $38$ SPB stars as a whole, the overall highest-$f_{\rm sv}$ regression model is the one delivered by prewhitening strategy $3$, as will be discussed further below.
The best regression model and other properties for all individual stars in our sample are discussed in Appendix~B.

\subsection{High-$f_{\rm sv}$ stars}\label{subsec:high_s_f_v_stars}

\begin{figure}
    \centering
    \resizebox{\hsize}{!}{\includegraphics{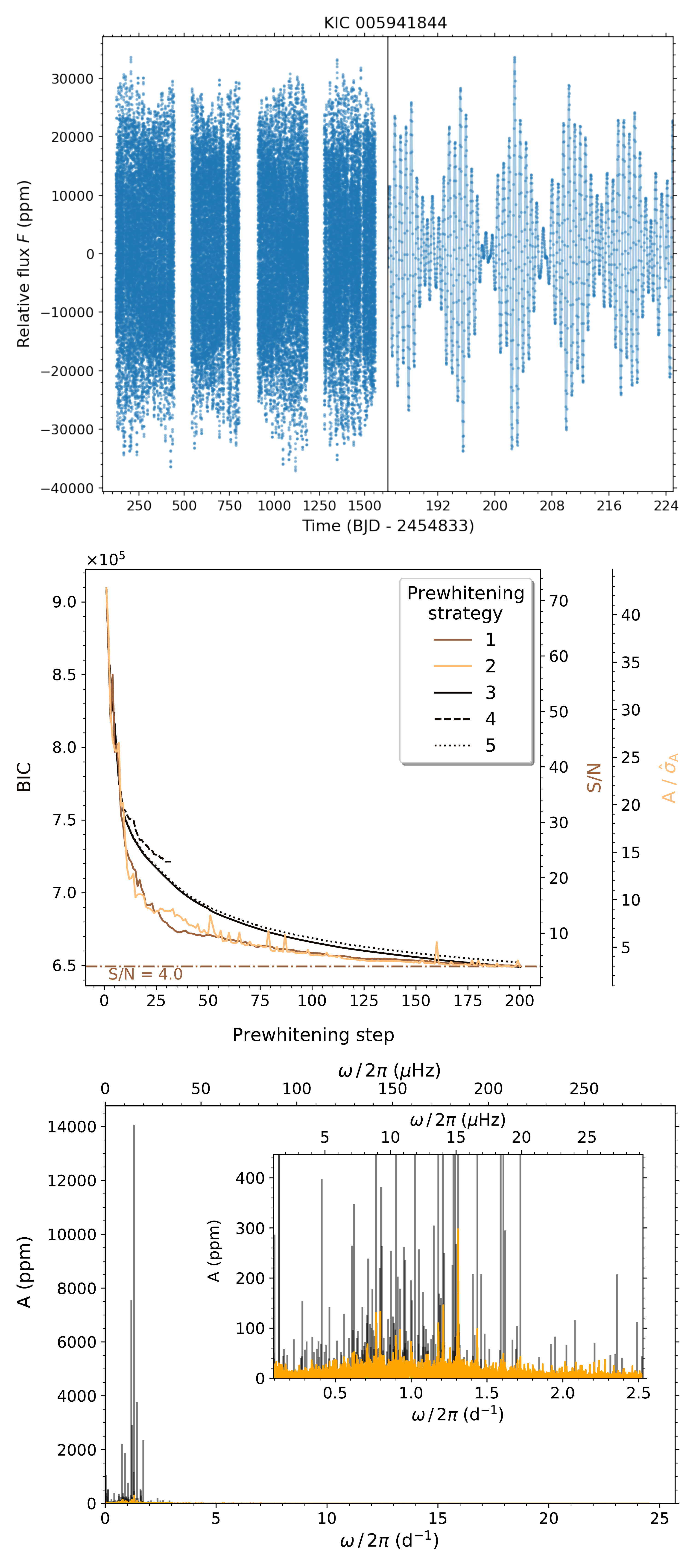}}
    \caption{\textit{Top:} Full light curve of the high-$f_{\rm sv}$ star KIC005941844 (left), and a small part of the light curve indicating its multi-periodic constant-frequency nature (right).
    \textit{Middle:} Comparison of the different stop criteria used for the five prewhitening strategies throughout the iterative prewhitening procedure, applied to the light curve of KIC005941844. The BIC values are relevant for strategies 3, 4, and 5, whereas the S/N value and the estimated amplitude $A$ expressed in estimated standard deviations $\hat{\sigma}_{A}$ are relevant for strategies 1 and 2, respectively.
    \textit{Bottom:} Lomb-Scargle periodogram (in orange) obtained after subtracting the light curve model obtained by prewhitening strategy 3 for KIC005941844, from the original light curve. The extracted frequencies for this model are displayed as grey lines with height equal to the extracted amplitude. Note the different y-axis scale for the inset compared to the main panel.
    }
    \label{fig:star_figure_KIC005941844}
\end{figure}

We discuss the light curve regression model selection process for high-$f_{\rm sv}$ stars guided by the example of KIC005941844, classified as an SPB star by \citet{2012AJ....143..101M} and analysed by \citet{may_phd_thesis} and \citet{2021NatAs.tmp...80P}. Its light curve is well explained by sinusoids of constant amplitude and frequency, as demonstrated by its high $f_{\rm sv}$ values and the light curve in the top panel of Fig.~\ref{fig:star_figure_KIC005941844}. The light curves of high-$f_{\rm sv}$ stars are, in general, (nearly) sinusoidal, because a large degree of variance is explained by the model generated by Eqn.~(\ref{eq:sum_sines}). The most prominent data gaps in the light curve of KIC005941844 are caused by the loss of {\em Kepler}'s CCD module 3, one year into the mission.

The middle panel of Fig.~\ref{fig:star_figure_KIC005941844} shows a comparison of the values of the metrics important for the stop criteria of all five strategies, throughout the iterative prewhitening process.
The values of these metrics as a function of the prewhitening step number indicate the progress of that process before frequency filtering is carried out, and are henceforth referred to as the `iteration progress curves'. If the gradient of the Bayesian information criterion (BIC) iteration progress curve (strategy 3, 4, or 5) converges to a value of zero, the regression model is approaching the optimal model according to the likelihood ratio test (LRT). The LRT compares the nested light curve regression models of this and the previous step based on the likelihoods inferred from the BIC values. A significant upward trend in the BIC iteration progress curve increases the p-value of the LRT to a value greater than the stop criterion, with the significance of that trend depending on the difference in BIC values and number of optimisation parameters of the nested regression models. The S/N iteration progress curve (strategy 1) represents the signal-to-noise ratio of the frequency extracted during each iterative prewhitening step. If it falls below the pre-set limit $S/N=4.0$ \citep{1993A&A...271..482B}, the stop criterion is triggered and iterative prewhitening stops. Finally, the $A/\hat{\sigma}_{A}$ iteration progress curve represents the amplitude of the frequency extracted during each iterative prewhitening step, in terms of its standard deviation. Iterative prewhitening for strategy 2 is stopped if the amplitude of the frequency extracted is consistent with $0.0$ ppm at $95$\% confidence level (i.e.\ $A/\hat{\sigma}_{A} \leq 1.96$).

The iteration progress curves provide a metric that dictates whether a light curve regression model is nearing its optimal parameters. This needs to be investigated on a star-to-star basis, because it intrinsically depends on the oscillation mode density in the (residual) LS periodograms and amplitudes attained by any of the periodic signals in the light curves. For KIC005941844, the iteration progress curves indicate that for all five iterative prewhitening strategies, the optimal or near-optimal model has been obtained, before any additional post-processing frequency filtering. The frequency filtering step removes unresolved frequencies and frequencies that do not have significant amplitudes (at a $95$\% confidence level), so that the final light curve regression models are obtained. The numbers of frequencies retained in the regression models after this frequency filtering step (i.e.\ the numbers of final resolved significant extracted frequencies) are indicated in Fig.~\ref{fig:s_f_v_comparison_prewhitening_strategies_SPB_stars} by the $n_e$ parameter for each strategy applied to the light curve of each star. 

\begin{figure*}
    \centering
    \includegraphics[width=18cm]{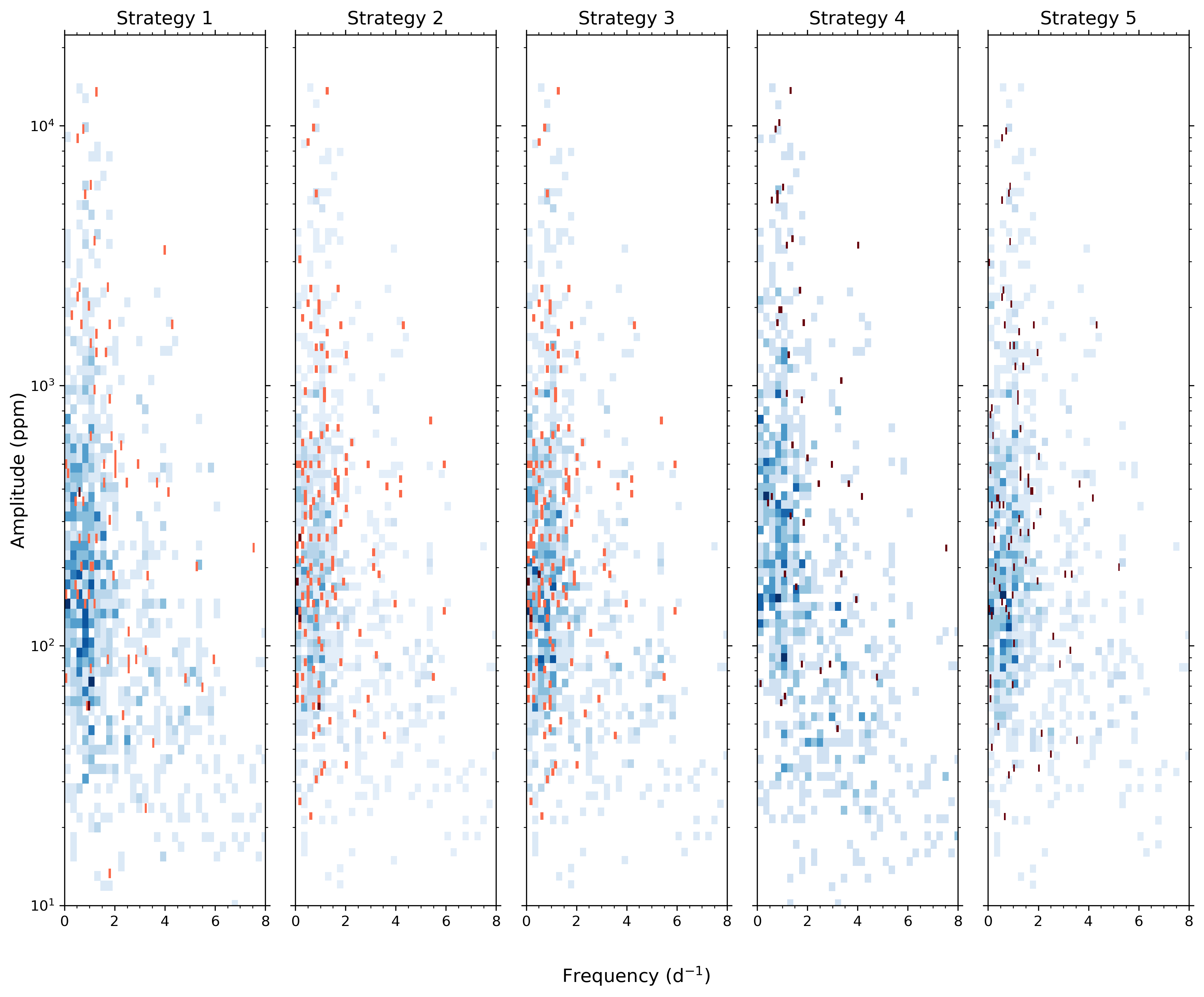}
    \caption{Distribution of the amplitudes and frequencies of (independent) signals not identified as combination frequencies for all 38 SPB stars in our sample, for each of the prewhitening strategies defined in Table~\ref{tab:strategy_table}. The colour indicates if these signals are involved in a candidate resonance (`red') or not (`blue'), and its gradient indicates the number of detected frequencies: colours are brighter if more signals are detected within a specific amplitude-frequency bin. The `red and blue distributions' are plotted independently, so that overlap between these distributions is the cause for the appearance of darker, maroon-like coloured bins belonging to the `red distribution'. The maximal frequency displayed in each panel is $8$ d$^{-1}$, which is the highest frequency of any of the detected (robust two-signal) candidate resonances (see Section~\ref{sec:candidate_resonance_search}, and especially, Fig.~\ref{fig:nonlinear_characteristics_two_largest_amplitudes}). We display independent signals with an amplitude $> 10$ ppm because very few independent signals have amplitudes smaller than this limit.}
    \label{fig:distribution_plot_scarcity_strategies}
\end{figure*}

The differences among the $f_{\rm sv}$ values of the regression models for all five strategies are small for the high-$f_{\rm sv}$ stars, as can be noticed when comparing these values listed for KIC005941844 in Fig.~\ref{fig:s_f_v_comparison_prewhitening_strategies_SPB_stars}.
Yet, the numbers of extracted frequencies, $n_{\rm e}$,  differ strongly. Good correspondence with respect to the number of extracted frequencies for strategies 2 and 3 occurs for almost all high-$f_{\rm sv}$ stars. For strategies 1 and 4, the parameter hinting selects the signal in the residual LS periodogram with the highest S/N value. Because we analyse $g$-mode oscillators, the parameter hinting for these strategies renders them prone to select a new frequency in a mode-scarce region of the periodogram, which is illustrated in Fig.~\ref{fig:distribution_plot_scarcity_strategies}, with the mode-scarce regions typically found at higher frequencies. Signals in mode-scarce regions of a typical LS periodogram of a $g$-mode oscillator are of lower amplitude than those in mode-dense regions. These signals therefore contribute less to $f_{\rm sv}$, explaining why for many high-$f_{\rm sv}$ stars, including KIC005941844, the stop criterion is triggered earlier for strategy 4 than for the other strategies. Conversely, the parameter hinting for strategies 2, 3, and 5 selects the largest-amplitude signal, and thus targets the mode-dense regions of such a typical periodogram. This hence reiterates the important role that mode density plays in frequency extraction, even for {\em Kepler} light curves, which offer the most optimal frequency resolution for space photometry available today.
Moreover, because different strategies focus their extraction efforts in different frequency regimes of the periodogram, the physical information extracted from these differing light curve regression models is different and can be complementary. 

The effect of a high mode density in the periodogram of KIC005941844 is clearly visible in the bottom panel of Fig.~\ref{fig:star_figure_KIC005941844}: dozens of small-amplitude signals persist in the residual LS periodogram, which are unresolved with respect to the \citetalias{1978Ap&SS..56..285L} criterion. Such unresolved signals can cause modulation of amplitudes and frequencies because they can interfere with other modes \citep[e.g.,][]{2016MNRAS.460.1970B} or be residual artefacts from imperfect prewhitening under the assumption that all oscillations have constant frequencies, amplitudes and phases. In this case the amplitudes of the extracted frequencies are large compared to the amplitudes of these unresolved signals, hence $f_{\rm sv}$ values are not strongly affected by such effects and remain high.

A sometimes neglected point in frequency analysis is that some of the free parameters in the fit to the light curve covary with other free parameters in the fit. Particularly strong covariances exist between the extracted frequencies and their corresponding phases.
As an example, neglecting such frequency-phase covariances leads to differences in the frequency values of the two largest-amplitude extracted frequencies of KIC005941844 at $2$-$\sigma$ level when comparing their amplitudes for the same mode frequencies extracted with strategy 5. Covariances among regression parameters not only influence the estimation of relative phases that are crucial to identifying candidate resonances, but can also exert influence on asteroseismic modelling relying on the frequency values (Bowman \& Michielsen, submitted).

From a purely mathematical point of view, parsimony dictates that any nested regression model that explains the same scaled fraction of variance as a more complex model (i.e.\ from the same model strategy but containing more free parameters) should be preferred over that more complex model, because of bias-variance trade-off \citep{claeskens_hjort_2008}. For this reason, one can use information criteria for model selection when the regression models are nested, as per each model strategy separately. However, one cannot use such criteria to distinguish between the capacity of the different regression models based on different stopping criteria and optimisation methods. Moreover, the aim of any asteroseismic data analysis is to deduce as many significant and resolved oscillation frequencies as possible (i.e.\ we wish to maximise 
$n_{\rm e}$ as listed in 
Fig.~\ref{fig:s_f_v_comparison_prewhitening_strategies_SPB_stars}). Since 
$n_{\rm e}$ is a discrete unknown that we want to maximise, within the adopted conditions on frequency resolution and significance of the modes' amplitudes, one should not use classical statistical model selection criteria to choose among the strategies 1 to 5. The case of KIC005941844 illustrates this: its five $f_{\rm sv}$ values differ by less than $\sim 6 \cdot 10^{-3}$ but the different prewhitening strategies lead to vastly different $n_{\rm e}$ values. A classical model selection procedure via a BIC or LRT applied to the models based on the different strategies by just using their $f_{\rm sv}$ alone would fail, as it would favour the model obtained by strategy 4 for KIC005941844. However, it is clear that the other four regression models with the vastly larger numbers of extracted modes capture more astrophysical information on, for example, 
the number of excited modes, 
potential mode interactions or rotational multiplets. Hence, the aim of finding a good regression model (high $f_{\rm sv}$) along with a maximum number of resolved oscillation frequencies of significant amplitude (high $n_{\rm e}$) favour strategies 3, 4 or 5 for this star. These three  regression models preferentially select signals at low frequencies, as shown in Fig.~\ref{fig:distribution_plot_scarcity_strategies}. The interplay between the physical and (purely) mathematical considerations in selecting the optimal iterative prewhitening model is challenging. For the SPB stars with high $f_{\rm sv}$, Fig.\,\ref{fig:explained_scaled_variance_comparison} shows all five prewhitening strategies to be almost equivalent by just looking at the summed $f_{\rm sv}$ for the 19 stars. However, adding the quest to maximise the number of resolved modes gives preference to strategies 2 or 3.
In any case, KIC005941844's large-amplitude oscillations and small-amplitude signals unresolved with respect to the \citetalias{1978Ap&SS..56..285L} resolution, as well as its high $f_{\rm sv}$ values, make it a prime target for nonlinear asteroseismic analysis.

\subsection{Low-$f_{\rm sv}$ stars}\label{subsec:low_s_f_v_stars}

\begin{figure}
    \centering
    \resizebox{\hsize}{!}{\includegraphics{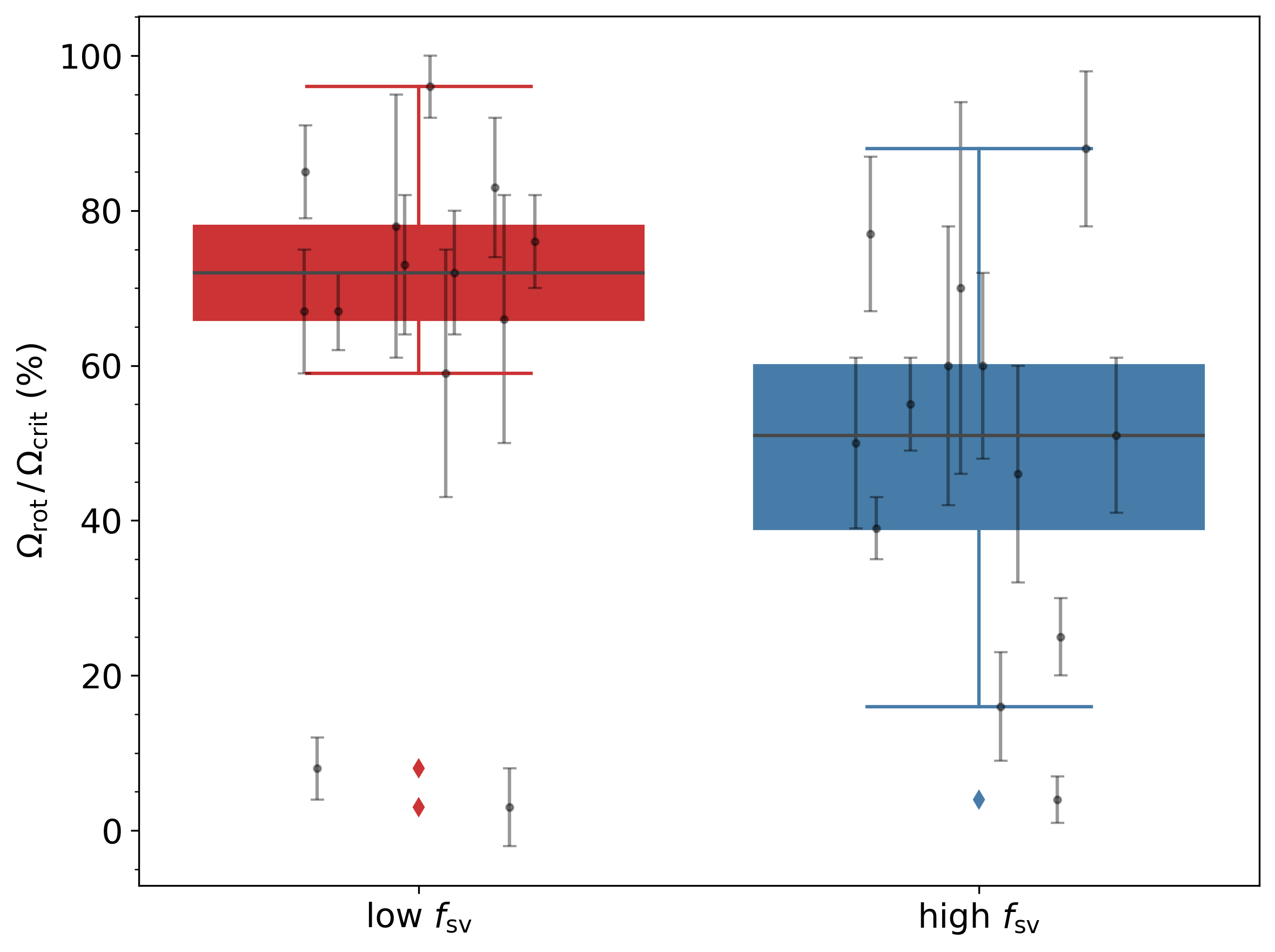}}
    \caption{Box-and-whisker plot of the rotation rates of 26 SPB stars obtained from averaging over 8 theories of core-boundary and envelope mixing by \citet{2021NatAs.tmp...80P}. These are grouped into the two classes defined in this work: low and high $f_{\rm sv}$, as indicated schematically on the x-axis. Whiskers are extended past the quartiles by maximally $1.5$ times the interquartile range. A single outlier for the low $f_{\rm sv}$ class is indicated with a rhombus mark. Individual rotation rates and estimated standard deviations are indicated in grey. The horizontal offset for these rotation rates was imposed for reasons of visibility without any additional meaning.}
    \label{fig:boxplot_rotation_correlation_classes}
\end{figure}

High mode densities are observed in the LS periodograms of all low-$f_{\rm sv}$ stars, because dozens of potential signals are left unresolved with respect to the \citetalias{1978Ap&SS..56..285L} resolution. Six of the nineteen low-$f_{\rm sv}$ stars furthermore display outburst-like features: marked departures from the baseline brightness of the original light curve, which are in phase with departures from the baseline in the residual light curve. We make a specific distinction between those stars that display these features in their light curves, and stars that do not. The former are referred to as `outbursting stars', whereas the latter are called `high mode density stars'. The discussion of the light curve regression model selection for low-$f_{\rm sv}$ stars is therefore guided by examples from both types of low-$f_{\rm sv}$ stars and the overall performance of the regression models in Fig.\,\ref{fig:explained_scaled_variance_comparison} is split up accordingly.

Oscillation mode density is expected to increase with increasing rotation rate due to rotational frequency shifts. No statistically significant difference is however found between the rotation rates of 26 of the high- and low-$f_{\rm sv}$ stars, derived by \citet{2021NatAs.tmp...80P}, as is illustrated by the box plot in Fig.~\ref{fig:boxplot_rotation_correlation_classes}. Weighted averages, on the other hand, indicate a difference, as illustrated in Tables \ref{tab:optimal_high_s_f_v} and \ref{tab:optimal_low_s_f_v}. The rotation rate for the high-$f_{\rm sv}$ star we discussed in the previous subsection is located around the centre of the respective class rotation rate distributions, whereas the rotation rates of the the low-$f_{\rm sv}$ stars we discuss in this subsection are at the lower boundaries of their distributions (see Tables \ref{tab:optimal_high_s_f_v} and \ref{tab:optimal_low_s_f_v} for the explicit values). The rotation rate of the high mode density star example, KIC008255796, is in fact an outlier in the rotation rate box-and-whisker plot for the low-$f_{\rm sv}$ class, next to the slowly-rotating SPB KIC008459899, which is suspected to be a double lined spectroscopic binary due to slightly larger O-C residuals of the fit to its observed spectrum \citep{2011A&A...526A.124L}. The latter star is the only (suspected) binary within the pseudoclass of low-$f_{\rm sv}$ stars. Whether and how its binary nature is correlated with its low rotation rate and low $f_{\rm sv}$ is unknown. With the exception of this binary and KIC008255796, the low $f_{\rm sv}$ values of the low-$f_{\rm sv}$ stars thus seem to be connected to rapid rotation.

\begin{table}
\centering
\caption{Rotation rates obtained from averaging over eight theories of core-boundary and envelope mixing by \citet{2021NatAs.tmp...80P} for the 19 high-$f_{\rm sv}$ stars, if applicable. The prewhitening strategy that delivers the highest-$f_{\rm sv}$ light curve regression model among the models generated by all five strategies, is noted in the `Strategy' column. The (approximate) amplitude of the largest-amplitude unresolved signal in the LS periodogram of the light curve subtracted by the highest-$f_{\rm sv}$ regression model is noted in the $A_{\rm unres/res}$ column, where it is symbolically divided by the amplitude of the largest-amplitude resolved signal in the LS periodogram of the original light curve.}
\label{tab:optimal_high_s_f_v}
\begin{tabular}{lrrr}
\toprule
KIC & $\Omega_{\rm rot} / \Omega_{\rm crit}$ (\%) & Strategy & $A_{\rm unres/res}$ (ppm)  \\
\midrule
$003459297$ & $55\pm 6$ & $2$ & $\sim 135/4600$ \\
$003839930$ & / & $2$ & $\sim330/10000$ \\
$003865742$ & $88\pm 10$ & $2$ & $\sim160/8750$ \\
$\mathbf{004930889}$ & $60\pm 12$ & $3$ & $\sim310/8000$ \\
$004936089$ & $16\pm 7$ & $3$ &  $\sim80/5700$ \\
$005084439$ & / & $2$ & $\sim175/3650$ \\
$005309849$ & $46\pm 14$ & $3$ & $\sim65/3400$ \\
$005941844$ & $39\pm 4$ & $3$ & $\sim300/14000$ \\
$\mathbf{006352430}$ & $50\pm11$ & $2$ & $\sim350/7400$ \\
$006780397$ & $70\pm 24$ &  $3$ & $\sim26/440$ \\
$007760680$ & $25\pm 5$ & $5$ & $\sim390/9800$  \\
$008324482$ & / & $2$ & $\sim175/2000$ \\
$008714886$ & $60\pm 18$ & $2$ &  $\sim155/2150$ \\
$008766405$ & $77\pm 10$ & $3$ & $\sim110/1750$ \\
$009020774$ & $51\pm 10$ & $2$ & $\sim75/7700$ \\
$009227988$ & / & $3$ & $\sim170/2200$ \\
$010220209$ & / & $2$ & $\sim44/1300$ \\
$010526294$ & $4\pm 3$ & $3$ & $\sim2000/14500$ \\
$010658302$ & / & $3$ & $\sim620/3000$ \\
\midrule
Average & $29$\,\tablefootmark{a} & N.A. & $300/5807$ \\
\bottomrule
\end{tabular}
\tablefoot{
Known binary stars \citep{2013A&A...553A.127P,2017A&A...598A..74P} have their KIC identifiers marked in boldface.\\
\tablefoottext{a}{Average weighted rotation rate, where the inverse variances ($1/\sigma^{2}_{\Omega_{\rm rot} / \Omega_{\rm crit}}$) are used as weights.}
}
\end{table}

\begin{table}
\centering
\caption{Same as Table~\ref{tab:optimal_high_s_f_v}, but for the 19 low-$f_{\rm sv}$ stars in our sample.}
\label{tab:optimal_low_s_f_v}
\begin{tabular}{lrrr}
\toprule
KIC & $\Omega_{\rm rot} / \Omega_{\rm crit}$ (\%) & Strategy  & $A_{\rm unres/res}$ (ppm) \\
\midrule
$003240411$ & $67\pm5$ & $3$ & $\sim 500/570$ \\
$003756031$ & / & $3$ & $\sim130/690$ \\
$004939281$\tablefootmark{*} & $78\pm17$ & $3$  & $\sim350/1850$ \\
$006462033$ & $85\pm 6$ & $2$ & $\sim60/600$ \\
$007630417$ & $66\pm 16$ & $2$ & $\sim160/540$ \\
$008057661$ & $59\pm 16$ & $3$  & $\sim50/310$ \\
$008087269$ & / & $3$ & $\sim175/1150$ \\
$008255796$ & $3\pm 5$ & $3$  & $\sim280/970$ \\
$008381949$ & $76\pm 6$ & $3$ & $\sim135/650$ \\
$\mathbf{008459899}$ & $8\pm 4$ & $2$ & $\sim145/640$ \\
$009715425$\tablefootmark{*} & $83\pm 9$ & $1$ & $\sim2400/4700$ \\
$009964614$ & / & $2$  & $\sim130/520$ \\
$010285114$\tablefootmark{*} & / & $3$  & $\sim380/730$ \\
$010536147$ & $67\pm 8$ & $3$ & $\sim145/2075$ \\
$011152422$\tablefootmark{*} & / & $3$ & $\sim135/540$ \\
$011360704$\tablefootmark{*} & $96\pm 4$ & $3$ & $\sim1225/2075$  \\
$011454304$ & / &  $3$ & $\sim350/535$ \\
$011971405$\tablefootmark{*} & $72\pm 8$ & $2$ & $\sim950/3500$  \\
$012258330$ & $73\pm 9$ & $5$ & $\sim155/500$ \\
\midrule
Average & $57$ & N.A. & $441/1218$ \\
\hspace{0.2cm} - Outb.\,\tablefootmark{a} & $90$ & N.A. & $907/2232$ \\
\hspace{0.2cm} - H.m.d.\,\tablefootmark{b}  & $44$ & N.A. & $186/750$ \\
\bottomrule
\end{tabular}
\tablefoot{
A suspected binary star \citep{2011A&A...526A.124L} has its KIC identifier marked in bold. The weighted averages for the rotation rates are computed in the same way as in Table~\ref{tab:optimal_high_s_f_v}. The numbers in the `Strategy' column denote the strategy for which the highest $f_{\rm sv}$ value is obtained.\\
\tablefoottext{*}{Outbursting star.}
\tablefoottext{a}{Average sample properties for the outbursting stars.}
\tablefoottext{b}{Average sample properties for the high mode density stars.}
}
\end{table}

\subsubsection{High mode density stars}\label{subsubsec:high_mode_density_stars}

\begin{figure}
    \centering
    \resizebox{\hsize}{!}{\includegraphics{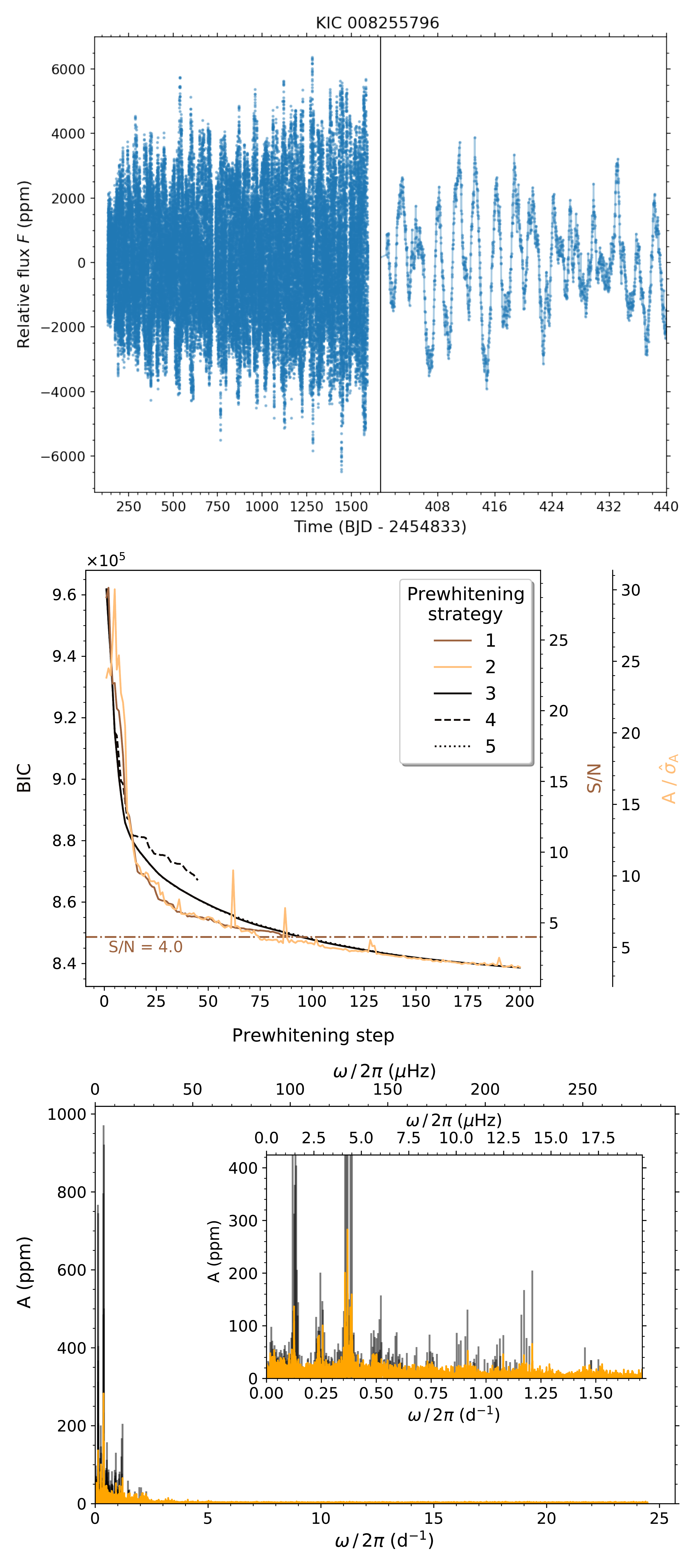}}
    \caption{Same as Fig.~\ref{fig:star_figure_KIC005941844}, but for high mode density star KIC008255796. 
    }
    \label{fig:star_figure_KIC008255796}
\end{figure}

\begin{figure}
    \centering
    \resizebox{\hsize}{!}{\includegraphics{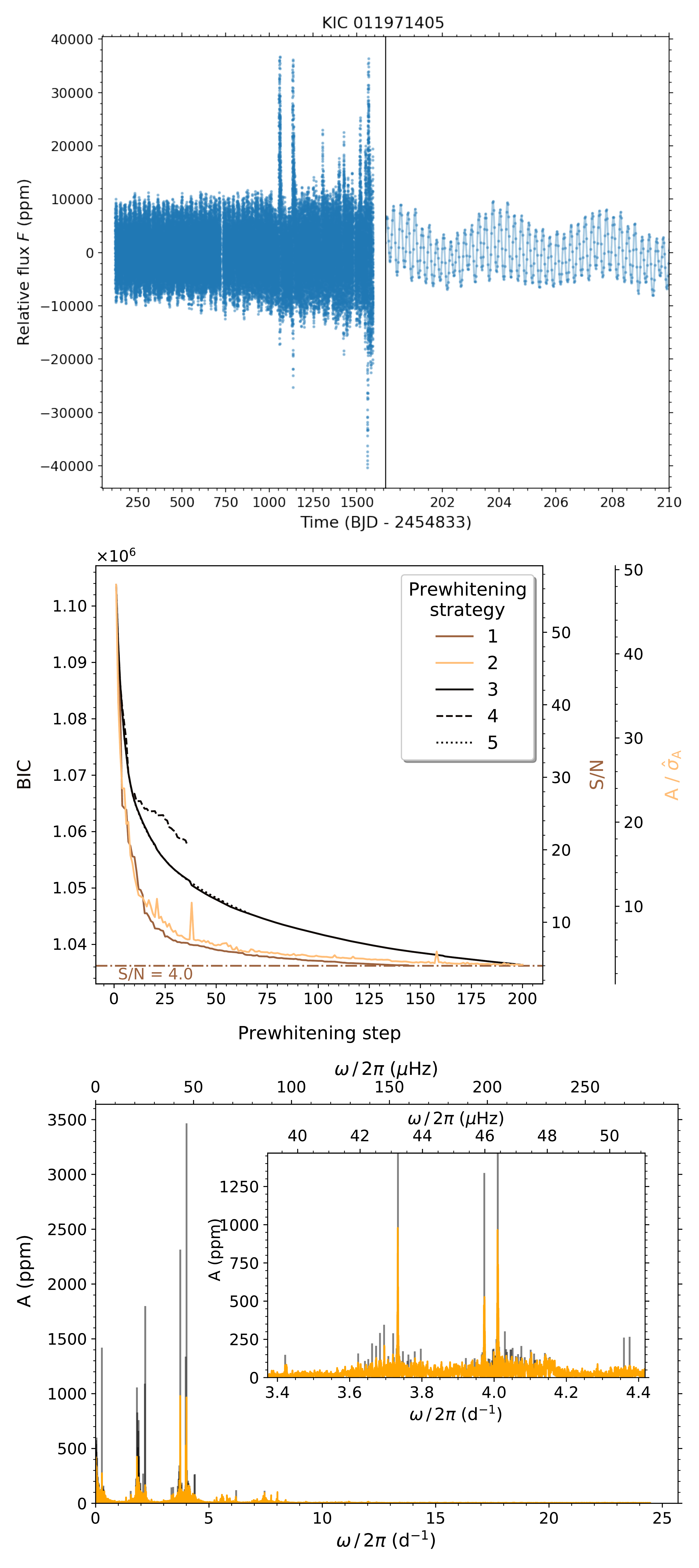}}
    \caption{Same as Fig.~\ref{fig:star_figure_KIC005941844}, but for outbursting star KIC011971405.}
    \label{fig:star_figure_KIC011971405}
\end{figure}

An example of a high mode density star, a low-$f_{\rm sv}$ star that does not display outburst-like features in its light curve, is KIC008255796. It was first identified as a misclassified B star by \citet{2018ApJ...854..168Z}. Its light curve is shown in the top panel of Fig.~\ref{fig:star_figure_KIC008255796} and displays periodic variations on longer time scales than KIC005941844, the high-$f_{\rm sv}$ star example. The characteristic properties of the light curves of high mode density stars such as KIC008255796 are not dissimilar to those of high-$f_{\rm sv}$ stars. The lower $f_{\rm sv}$ values attained by high mode density stars can be attributed to the fact that on average maximal excursions from the baseline brightness are smaller, so that small-amplitude unresolved signals have a larger impact on the $f_{\rm sv}$ value. The maximal excursion from the light curve baseline for KIC005941844 is $\sim 35$~ppt, whereas this value is only $\sim 6$~ppt for KIC008255796, as can be deduced from the top panels of Fig.~\ref{fig:star_figure_KIC005941844} and Fig.~\ref{fig:star_figure_KIC008255796}.

The iteration progress curves of high mode density stars are also not dissimilar to those of high-$f_{\rm sv}$ stars, as is illustrated by comparing the middle panels of Figs \ref{fig:star_figure_KIC005941844} and \ref{fig:star_figure_KIC008255796}, obtained for KIC005941844 and KIC008255796, respectively. The strategy 2 and 3 iteration progress curves of KIC008255796 indicate that more frequencies need to be extracted to reach the `optimal model' for that strategy. This is not unexpected, because these strategies focus their extraction efforts on high mode density regions of the periodogram, which are strongly present in the LS periodograms of high mode density stars. The other strategies have triggered their stop criteria, and those in which hinting is S/N-based are especially strict, as expected.

The eponymous high mode density within the periodogram is well-illustrated in the bottom panel of Fig.~\ref{fig:star_figure_KIC008255796} for KIC008255796. Higher mode densities cause more signals to be unresolved with respect to the \citetalias{1978Ap&SS..56..285L} criterion, and therefore increase the probability that a larger-amplitude signal is unresolved. Combined with the on average smaller amplitude signals observed in high mode density stars compared to high-$f_{\rm sv}$ stars, which are a corollary of the observed differences in maximal excursions from the light curve baseline, a larger influence of unresolved signals on $f_{\rm sv}$ is expected. Indeed, as can be derived from Tables \ref{tab:optimal_high_s_f_v} and \ref{tab:optimal_low_s_f_v}, the average unresolved/resolved amplitude for high mode density stars is $\sim 186/750$ ppm, whereas that value is $\sim 300/5807$ ppm for high-$f_{\rm sv}$ stars. 

The similarities noted in the light curve and Fourier properties of high mode density low-$f_{\rm sv}$ stars and high-$f_{\rm sv}$ stars render their classification somewhat artificial. This is why we refer to the high- and low-$f_{\rm sv}$ stars as pseudoclasses. It is however expected that amplitude modulation (and frequency modulation) more frequently occurs for high mode density stars, as they on average exhibit more unresolved signals. Whether that implies that the observed differences in $f_{\rm sv}$ between the two pseudoclasses are not of a purely instrumental origin (i.e.\ not being able to resolve high mode densities due to limited time span of the photometry), remains to be validated. What is clear, however, is that the amplitude modulation observed in some high mode density stars could originate from beating of close frequency pairs \citep{2016MNRAS.460.1970B}, or nonlinear mode interactions in the `intermediate' regime \citep[e.g.,][]{1997A&A...321..159B, 1998BaltA...7...21G}. Hence, some high mode density stars also are prime targets for in-depth nonlinear asteroseismic analysis.

\subsubsection{Outbursting stars}\label{subsubsec:outbursting_stars}

An example of a low-$f_{\rm sv}$ star that displays the eponymous outburst-like features in its light curve, an outbursting star, is KIC011971405.
It is considered to be a prototypical oscillating Be star by \citet{2015MNRAS.450.3015K}, and was first identified by \citet{2012AJ....143..101M} as an SPB oscillator. \citet{2017A&A...598A..74P} spectroscopically confirmed it as a Be star, and studied its variability and outbursts in detail. The days-long outburst-like features in its light curve are readily visible in the upper panel of Fig.~\ref{fig:star_figure_KIC011971405}. Similar features can be observed in the light curves of the other outbursting stars, with differing lengths and amplitudes. However, as illustrated in the same panel of Fig.~\ref{fig:star_figure_KIC011971405}, the parts of the light curve of KIC011971405 away from these outburst-like features are very similar to that of a non-outbursting SPB oscillator. This similarity is also noted for the other outbursting stars, and fits within the angular momentum transport model put forward by \citet{2020A&A...644A...9N} that is used to describe Be star outbursts. Within this model gravito-inertial modes stochastically excited by core convection dominate the LS periodogram
at the time of outbursts, whereas $g$ modes excited by the $\kappa$ mechanism prevail for the entire time span of the light curve. Hence, amplitude and frequency modulation is expected for all outbursting stars, which gives rise to high mode densities in the periodogram. For KIC011971405 modulation has been detected and characterised by \citet{2017A&A...598A..74P}. Interestingly, while for five of six outbursting stars these high mode densities are distinctly grouped, KIC011152422 displays a high mode density in its LS periodogram that spans from $0$ to $1$ d$^{-1}$.

The S/N stop criteria trigger an early stop to the prewhitening (i.e.\ before 200 frequencies are extracted) for almost all outbursting stars, an example of which can be observed for KIC01197405 in the middle panel of Fig.~\ref{fig:star_figure_KIC011971405}. High mode densities are the main cause for these early triggers. The exception is KIC010285114, whose outburst amplitude, in comparison to the typical excursions from the baseline in its light curve, is the smallest for all outbursting stars. The stop criteria of strategies 4 and 5 trigger early abortion of the prewhitening for all outbursting stars, whereas the iteration progress curves for strategies 2 and 3 indicate more frequencies can be extracted. 

The high mode densities in the LS periodograms and the outburst-like features in the light curves are also the prime factors to which the low $f_{\rm sv}$ values for outbursting stars can be attributed. Because of these high mode densities we do not expect any significant changes in $f_{\rm sv}$ after incorporating additional frequencies extracted by strategies 2 and 3 for KIC01197405, and the other outbursting stars. The observed amplitude modulation and outbursts in the outbursting stars in our sample could be linked to many $g$ modes coming into phase with each other, some of which by nonlinear interaction \citep{2015MNRAS.450.3015K}. If this is the sole origin of the outbursts, it corroborates the idea that oscillating Be stars are complex analogues of SPB stars \citep{2006ESASP.624E.131A}, and supports the idea of the artificiality of the high- and low-$f_{\rm sv}$ pseudoclasses that was discussed in the previous subsection. The \citet{2015MNRAS.450.3015K} theory relies upon a light curve model similar to the one utilised in this work, consisting of (interacting) sinusoids of constant amplitude and frequency. \citet{2015MNRAS.450.3015K} show that the bulk of the variability of KIC011971405 can be modelled in such a manner. Whether such a light curve model is sufficient to explain the bulk of the observed modulation for all of the outbursting B stars observed by {\it Kepler\/} remains to be verified, but this is outside of the scope of this work. 
As seen from Figs\,\ref{fig:s_f_v_comparison_prewhitening_strategies_SPB_stars}
and \ref{fig:explained_scaled_variance_comparison}, the choice in prewhitening strategy clearly exerts a stronger influence for outbursting stars than for the other SPB stars. 

\section{Results: candidate resonance characteristics}\label{sec:candidate_resonance_search}

We now discuss the results of our search for candidate nonlinear $g$-mode resonances within the light curve models obtained for the prewhitening strategies applied to all stars in our sample.
The degree to which nonlinear interactions are important increases with pulsation amplitude. We therefore limit our search for candidate resonances to the combination frequencies that include as a parent frequency at least one of the two largest-amplitude oscillation signals detected in the light curves, or both. We hereafter refer to such candidate resonances as `two-signal' candidate resonances. We also verify whether the number of two-signal candidate resonances is correlated to the binary or outbursting nature of the SPB stars, or their $f_{\rm sv}$ pseudo-classification.
More candidate resonances that do not involve these two largest-amplitude signals may be detectable. We discuss these additional candidate resonances for each star in the sample in Appendix~B, in which we also indicate the priority for nonlinear asteroseismic follow-up. 

\begin{table}[h!]
\centering
\caption{Number of two-signal candidate resonances for each of the five prewhitening strategies applied to all stars in our sample, and the number of two-signal candidate resonances defined as `Robust', i.e.\ fulfilling Eqn.~(\ref{eq:robust_match}).}
\label{tab:2_largest_amplitude_frequencies_count_overview_strategies}

\begin{tabular}{l|llllll}
\toprule
\multicolumn{1}{c|}{\multirow{2}{*}{KIC}} & \multicolumn{6}{c}{Prewhitening strategy}  \\
   &         1 &        2 &        3 &         4 &        5 & Robust \\
\midrule
$\mathbf{003240411}$ &   $2$/$3$ & $2$/$17$ & $\mathbf{1}$/$\mathbf{16}$ &   $0$/$0$ &  $0$/$0$ &  $0$/$0$ \\
$003459297$ &   $3$/$3$ & $\mathbf{3}$/$\mathbf{11}$ & $4$/$10$ &   $3$/$3$ &  $2$/$2$ &  $0$/$0$ \\
$\mathbf{003756031}$ &   $6$/$7$ & $5$/$20$ & $\mathbf{5}$/$\mathbf{20}$ &   $0$/$0$ & $2$/$13$ &  $1$/$6$ \\
$003839930$ &   $5$/$5$ & $\mathbf{4}$/$\mathbf{16}$ & $4$/$16$ &   $3$/$3$ &  $1$/$1$ &  $1$/$1$ \\
$003865742$ &   $4$/$7$ & $\mathbf{1}$/$\mathbf{14}$ & $1$/$13$ &   $4$/$4$ & $1$/$16$ &  $0$/$6$ \\
$004930889$ &   $4$/$6$ &  $2$/$9$ &  $\mathbf{2}$/$\mathbf{9}$ &   $1$/$1$ &  $1$/$6$ &  $0$/$4$ \\
$004936089$ &   $4$/$7$ &  $2$/$7$ &  $\mathbf{2}$/$\mathbf{7}$ &   $1$/$1$ &  $4$/$6$ &  $2$/$2$ \\
$\mathit{004939281}$ &   $1$/$2$ & $2$/$12$ & $\mathbf{2}$/$\mathbf{12}$ &   $0$/$0$ &  $0$/$4$ &  $0$/$0$ \\
$005084439$ &  $8$/$14$ & $\mathbf{9}$/$\mathbf{27}$ &  $9$/$27$ & $10$/$20$ &  $7$/$9$ &  $6$/$6$ \\
$005309849$ & $10$/$10$ &  $4$/$7$ &  $\mathbf{4}$/$\mathbf{7}$ &   $7$/$7$ &  $3$/$6$ &  $3$/$5$ \\
$005941844$ & $12$/$14$ & $9$/$18$ & $\mathbf{9}$/$\mathbf{18}$ &   $5$/$5$ & $5$/$15$ & $5$/$12$ \\
$006352430$ &   $1$/$1$ & $\mathbf{6}$/$\mathbf{16}$ & $6$/$16$ &   $0$/$0$ & $5$/$15$ & $5$/$12$ \\
$\mathbf{006462033}$ &   $2$/$2$ &  $\mathbf{2}$/$\mathbf{8}$ &  $2$/$8$ &   $0$/$0$ &  $1$/$1$ &  $0$/$0$ \\
$006780397$ &   $6$/$7$ & $2$/$12$ & $\mathbf{2}$/$\mathbf{12}$ &   $0$/$0$ & $0$/$10$ &  $0$/$8$ \\
$\mathbf{007630417}$ &   $2$/$3$ &  $\mathbf{2}$/$\mathbf{4}$ &  $2$/$4$ &   $3$/$3$ &  $3$/$5$ &  $2$/$4$ \\
$007760680$ &   $4$/$5$ &  $3$/$3$ &  $3$/$3$ &   $0$/$0$ &  $\mathbf{5}$/$\mathbf{8}$ &  $3$/$4$ \\
$\mathbf{008057661}$ &   $2$/$2$ &  $0$/$6$ &  $\mathbf{0}$/$\mathbf{6}$ &   $2$/$2$ &  $0$/$6$ &  $0$/$4$ \\
$\mathbf{008087269}$ &   $1$/$3$ &  $2$/$9$ &  $\mathbf{4}$/$\mathbf{9}$ &   $2$/$2$ &  $2$/$7$ &  $2$/$4$ \\
$\mathbf{008255796}$ &   $0$/$3$ & $0$/$17$ & $\mathbf{0}$/$\mathbf{17}$ &   $0$/$0$ & $0$/$14$ & $0$/$10$ \\
$008324482$ &   $6$/$6$ & $\mathbf{4}$/$\mathbf{13}$ & $4$/$13$ &   $2$/$2$ &  $3$/$4$ &  $2$/$3$ \\
$\mathbf{008381949}$ &   $1$/$1$ & $2$/$16$ & $\mathbf{2}$/$\mathbf{16}$ &   $0$/$0$ &  $0$/$2$ &  $0$/$2$ \\
$\mathbf{008459899}$ &   $1$/$4$ & $\mathbf{3}$/$\mathbf{29}$ & $3$/$29$ &   $0$/$0$ &  $1$/$3$ &  $1$/$2$ \\
$008714886$ &   $6$/$6$ & $\mathbf{6}$/$\mathbf{17}$ & $6$/$17$ &   $1$/$1$ & $3$/$11$ & $3$/$11$ \\
$008766405$ &   $5$/$5$ & $7$/$16$ & $\mathbf{7}$/$\mathbf{16}$ &   $5$/$5$ & $7$/$16$ & $6$/$12$ \\
$009020774$ &   $4$/$5$ & $\mathbf{4}$/$\mathbf{11}$ & $4$/$11$ &   $4$/$5$ &  $1$/$2$ &  $0$/$1$ \\
$009227988$ &   $3$/$4$ & $1$/$13$ & $\mathbf{1}$/$\mathbf{13}$ &   $3$/$4$ & $2$/$13$ &  $0$/$2$ \\
$\mathit{009715425}$ &   $\mathbf{3}$/$\mathbf{5}$ &  $1$/$4$ &  $1$/$4$ &   $1$/$1$ &  $2$/$7$ &  $1$/$1$ \\
$\mathbf{009964614}$ &   $1$/$3$ & $\mathbf{0}$/$\mathbf{13}$ & $0$/$13$ &   $0$/$0$ & $0$/$10$ &  $0$/$2$ \\
$010220209$ &   $2$/$5$ &  $\mathbf{2}$/$\mathbf{8}$ &  $2$/$8$ &   $1$/$9$ &  $0$/$0$ &  $0$/$0$ \\
$\mathit{010285114}$ &   $6$/$6$ & $4$/$10$ & $\mathbf{4}$/$\mathbf{10}$ &   $0$/$0$ &  $0$/$1$ &  $1$/$1$ \\
$010526294$ &   $7$/$8$ &  $7$/$8$ &  $\mathbf{7}$/$\mathbf{8}$ &   $4$/$4$ & $9$/$10$ &  $7$/$8$ \\
$\mathbf{010536147}$ &   $0$/$1$ &  $1$/$7$ &  $\mathbf{1}$/$\mathbf{7}$ &   $0$/$0$ & $1$/$11$ &  $1$/$7$ \\
$010658302$ &   $4$/$4$ &  $3$/$9$ & $\mathbf{4}$/$\mathbf{17}$ &   $1$/$1$ &  $3$/$3$ &  $1$/$1$ \\
$\mathit{011152422}$ &   $3$/$3$ & $2$/$15$ & $\mathbf{2}$/$\mathbf{16}$ &   $3$/$6$ & $2$/$13$ &  $2$/$4$ \\
$\mathit{011360704}$ &   $6$/$7$ & $\mathbf{3}$/$\mathbf{19}$ & $3$/$19$ &   $2$/$2$ & $1$/$11$ &  $1$/$8$ \\
$\mathbf{011454304}$ &   $3$/$5$ & $3$/$11$ & $\mathbf{3}$/$\mathbf{11}$ &   $0$/$0$ &  $2$/$2$ &  $1$/$1$ \\
$\mathit{011971405}$ &   $5$/$9$ & $\mathbf{3}$/$\mathbf{16}$ & $4$/$17$ &   $2$/$2$ &  $1$/$2$ &  $0$/$0$ \\
$\mathbf{012258330}$ &   $5$/$5$ & $4$/$10$ & $4$/$11$ &   $3$/$3$ &  $\mathbf{5}$/$\mathbf{8}$ &  $3$/$6$ \\
\bottomrule
\end{tabular}
\tablefoot{
Result format: S/N $\geq 5$ restriction imposed ~/~ no S/N restriction imposed. Boldface for the overall highest-$f_{\rm sv}$ strategy.\\
KIC number style: outbursting stars - italics, high mode density stars - boldface, high-$f_{\rm sv}$ stars - regular.
}
\end{table}

\begin{figure*}
    \centering
    \includegraphics[width=0.49\textwidth]{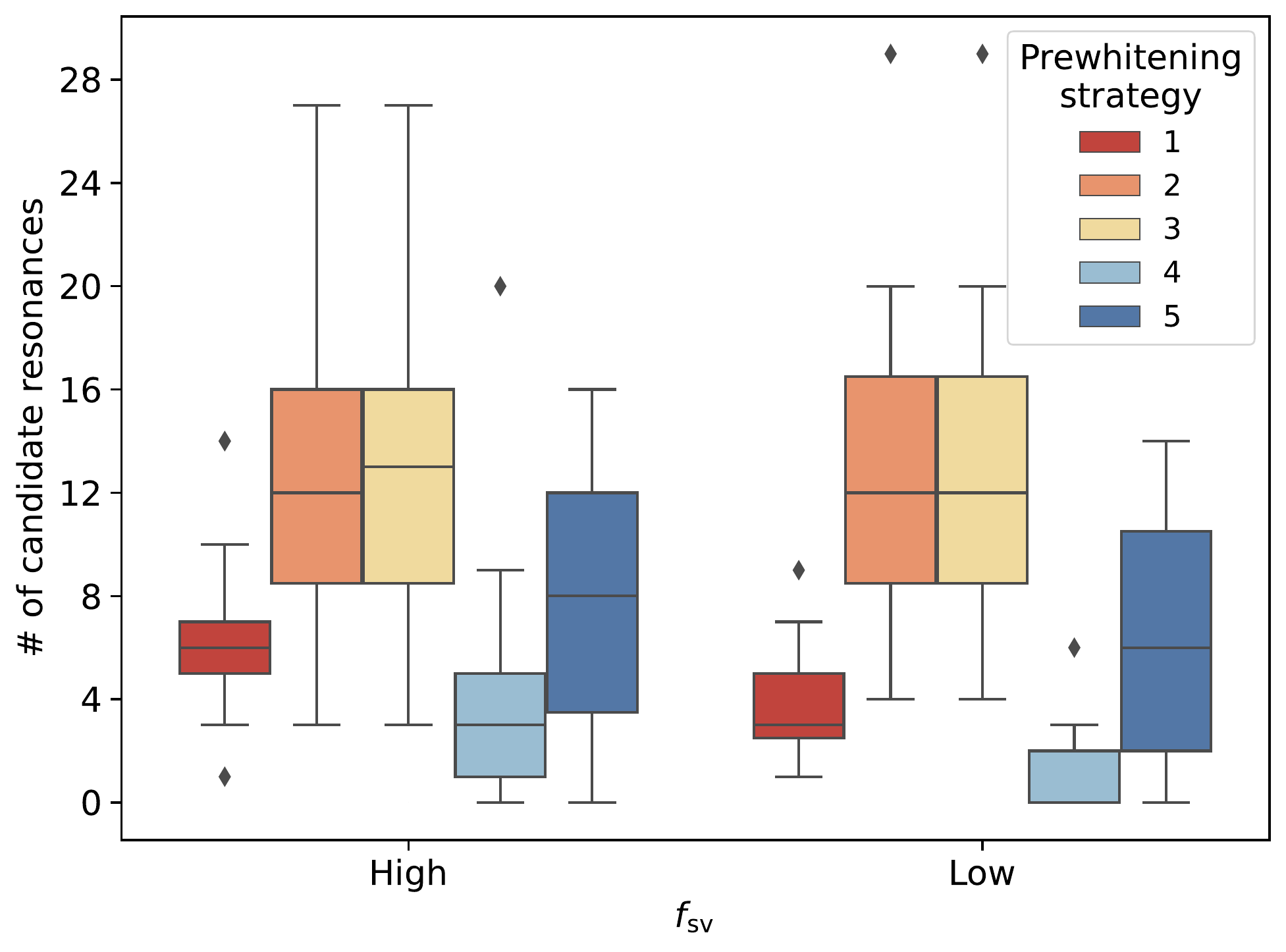}
    \includegraphics[width=0.49\textwidth]{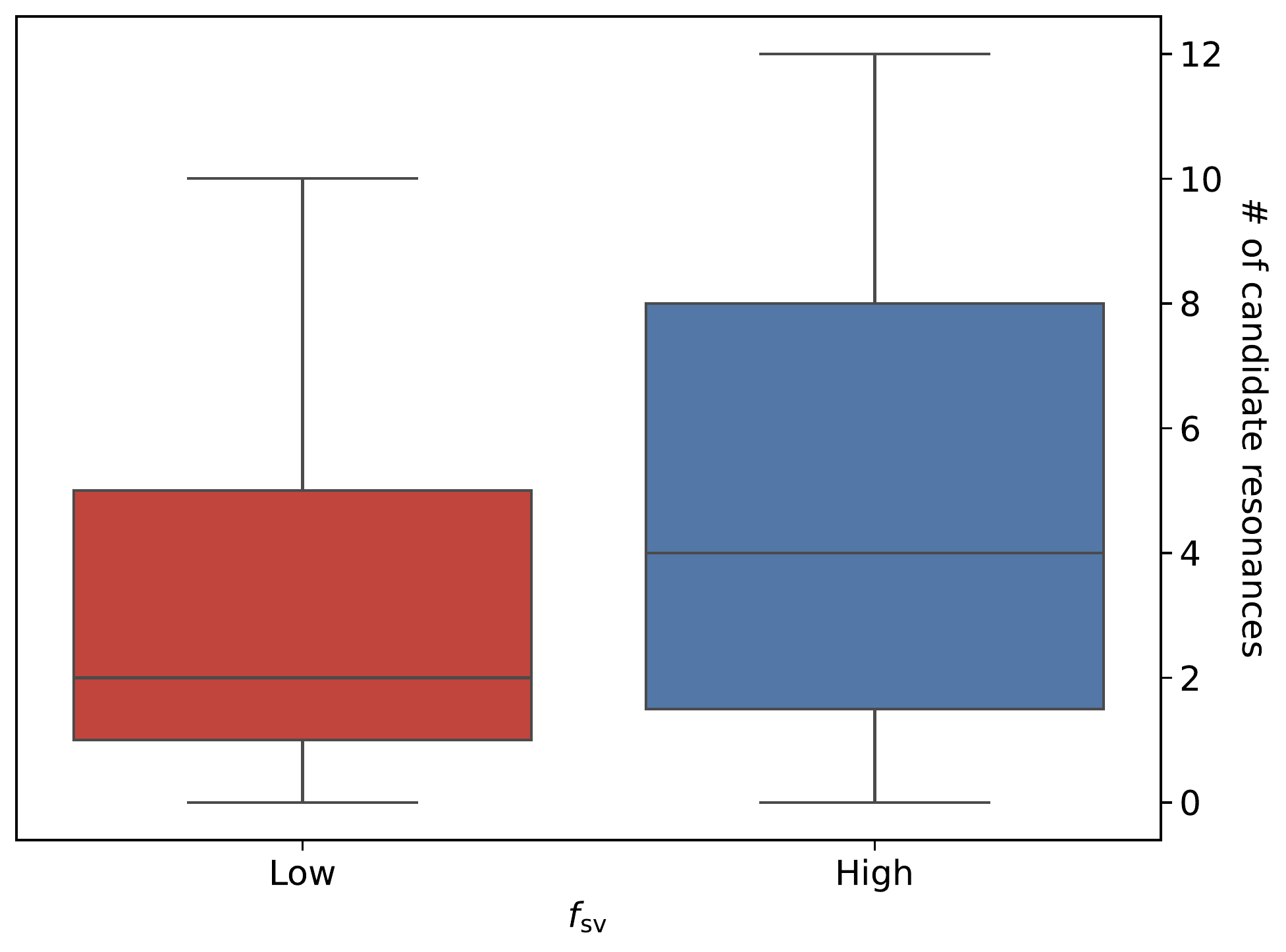}
    \caption{\textit{Left panel:} box-and-whisker plot of the number of identified two-signal candidate resonances, grouped into the two pseudoclasses defined in this work: low and high $f_{\rm sv}$. The colour depicts the prewhitening strategy employed to generate the light curve models. \textit{Right panel:} box-and-whisker plot of the number of identified robust two-signal candidate resonances (defined by Eqn.~(\ref{eq:robust_match})), grouped by pseudoclass. Whiskers and outliers for both panels are indicated in a similar way to how they are indicated in Fig.~\ref{fig:boxplot_rotation_correlation_classes}.}
    \label{fig:comparison_s_f_v_robust_resonance}
\end{figure*}

The strategy discussed in subsection \ref{subsec:candidate_resonant_oscillations} is employed to identify two-signal candidate resonances in the frequency-locked regime of mode interaction for the models generated by all five strategies for each star. 
To do so, we verify whether the highest-$f_{\rm sv}$ strategy that performs nonlinear least-squares optimisation of the frequencies (i.e.\ one of strategies $1$-$4$) contains unique matches in frequency to within 1 Rayleigh limit at the 1-$\sigma$ level, with the frequencies in the linear regression model generated by prewhitening strategy 5:
\begin{equation}
    \exists! \nu_{f_{\rm sv}}:\hspace{0.2cm}\left|\nu_{\rm lin} - \nu_{f_{\rm sv}}\right| - \sigma_{\nu_{\rm lin}} - \sigma_{\nu_{f_{\rm sv}}} \leq \mathfrak{R}_{\nu} \,.
    \label{eq:robust_match}
\end{equation}
In Eqn.~(\ref{eq:robust_match}) $\nu_{f_{\rm sv}}$ denotes the frequency of the model that attains the highest-$f_{\rm sv}$ among all strategies that employ nonlinear least-squares optimisation, and $\nu_{\rm lin}$ denotes the frequency obtained for prewhitening strategy $5$. We refer to candidate resonances for which Eqn.~(\ref{eq:robust_match}) is fulfilled as `robust' candidate resonances and discuss how their occurrence covaries with (pseudo)classification in subsection \ref{subsec:numbers_candidate_resonances}. We also discuss how this is reflected in the characteristic properties of these resonances in subsection \ref{subsubsec:characteristics_candidate_resonances}.

\citet{2015MNRAS.448L..16B} and \citet{2016A&A...585A..22Z} note that the \citet{1993A&A...271..482B} S/N detection threshold of $4.0$ is low as a detection threshold for signals in K2 data and {\em Kepler} short-cadence data, respectively. On the other hand, \citet{2009A&A...506..111D} and \citet{2009A&A...506..471D} conclude that not all signals failing the \citet{1993A&A...271..482B} threshold correspond to noise peaks (i.e.\ S/N-based methods are not clearcut in their optimal cut-off value as significance criterion). For the detection of resonances, we therefore make an additional distinction between frequencies with S/N $\geq 5$ and S/N $< 5$, when analysing candidate resonances in the following subsections. 

\subsection{Numbers of identified candidate resonances}\label{subsec:numbers_candidate_resonances}

The numbers of two-signal candidate resonances, with an indication of how many of those candidate resonances have S/N values $\geq 5$, are displayed in Table~\ref{tab:2_largest_amplitude_frequencies_count_overview_strategies} for all five strategies applied to each star in our sample. The number of robust two-signal candidate resonances and its S/N $\geq 5$ counterpart are also shown. It is apparent that for every star in our sample there is at least one prewhitening strategy that delivers a final light curve model in which two-signal candidate resonances are identified. For 32 out of 38 stars ($\sim 84$\%) we find robust resonance identifications in the definition in Eqn.~(\ref{eq:robust_match}). Imposing the additional S/N $\geq 5$ restriction lowers this number to 24 out of 38 stars ($\sim 63$\%). More candidate resonances are identified if we loosen the restriction on involvement of the two largest-amplitude candidate resonances. Candidate resonances are thus omnipresent in the light curves of the SPB stars we consider in this work, and SPB stars in general are therefore intrinsic nonlinear nonradial oscillators.

\begin{figure}
    \centering
    \resizebox{\hsize}{!}{\includegraphics{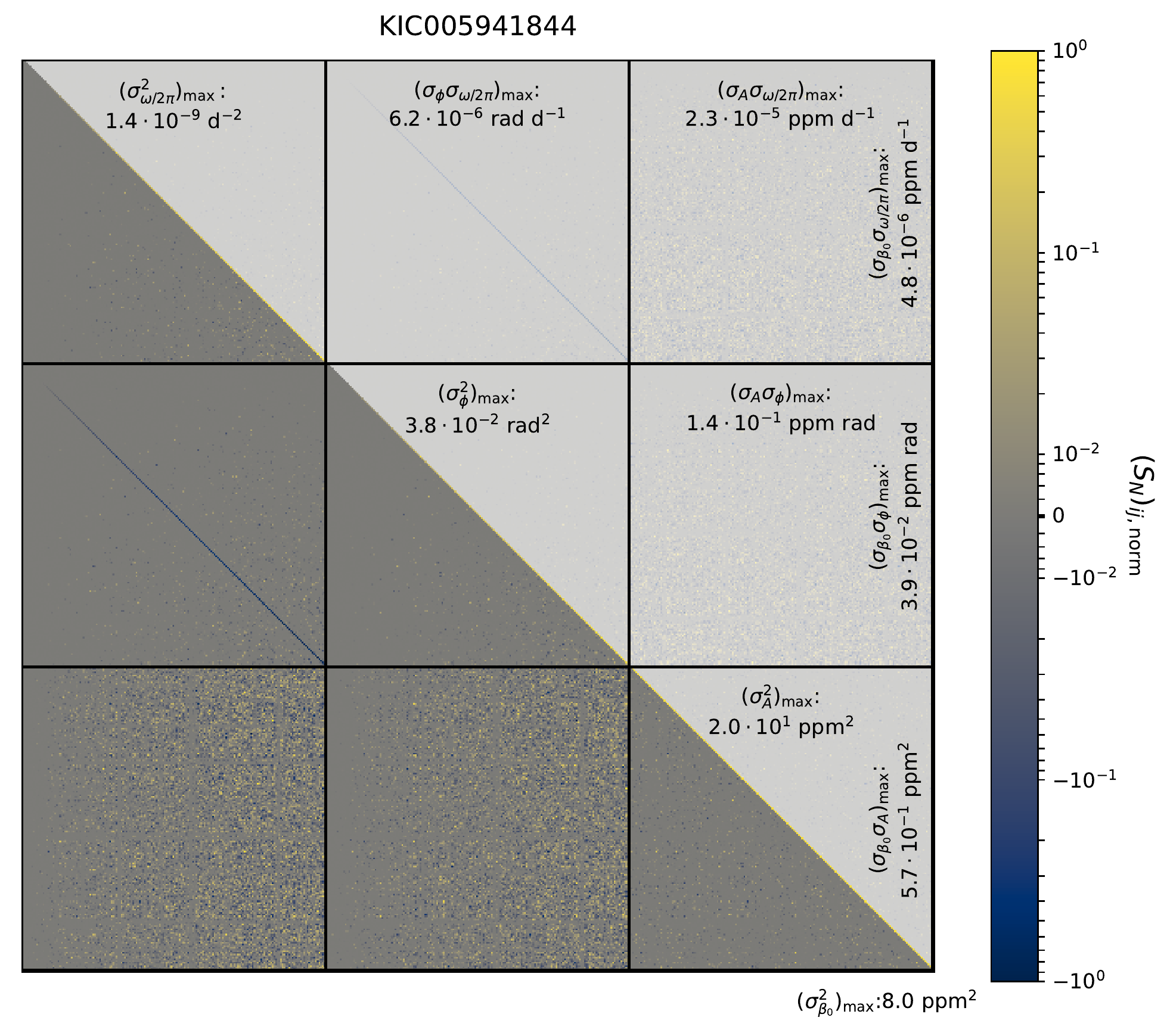}}
    \caption{Covariance matrix of the best regression model of the light curve of KIC005914844 (obtained from prewhitening strategy 3) in the block-normalised form described in Appendix A.4 and Fig.~A.1. The colour map shows the normalised (co)variance values ($S_N$)$_{ij,\,{\rm norm}}$ and the change in darkness between the upper and lower triangular part of the matrix is only for visualisation purposes.
    }
    \label{fig:covar_matrix_KIC005941844}
\end{figure}

The covariance matrices of the light curve models obtained from applying the prewhitening strategies to all stars in the sample vary from strategy to strategy and from star to star. The figures of Appendix A display visualisations of the covariance matrices obtained for the highest-$f_{\rm sv}$ prewhitening strategy (i.e.\ the best regression model), in a block-normalised form, which is discussed in detail in Appendix A.4.
An example of such a visualisation of a covariance matrix is given in Fig.~\ref{fig:covar_matrix_KIC005941844}, which displays the covariance matrix of the best regression model of the light curve of the high-$f_{\rm sv}$ star KIC005941844. Large numbers of weak inter-parameter covariations and the strong frequency-phase covariations mentioned in Sect.~\ref{subsec:high_s_f_v_stars} can be seen, similar to what is observed in all covariance matrices displayed in Appendix A.4. These covariations differ from star to star. Several covariance and correlation matrices display a structured pattern within their different blocks (i.e.\ they are structured) that indicates the strongest inter-parameter covariations and correlations.

The covariance matrix propagates into the relative frequencies and phases defined in Eqns~(\ref{eq:frequency_relation}) and (\ref{eq:relative_phase_relation}), which are used to identify a candidate resonance. The occurrence of robust candidate resonances as indicated in Table \ref{tab:2_largest_amplitude_frequencies_count_overview_strategies} thus connects to the covariance matrix structure among differing light curve models.

In asteroseismic forward modelling one compares observed mode frequencies with frequencies obtained from models of stellar evolution. As \citet{2016MNRAS.456.2183D} note for asteroseismic modelling based on solar-like oscillations, it is essential to account for regression parameter covariances and correlations in this process.
Regression parameter covariances and correlations can also impact asteroseismic forward modelling based on gravity modes, as discussed in Sect. \ref{subsec:high_s_f_v_stars}. These covariances and correlations also impact nonlinear asteroseismic modelling, which utilises extracted amplitudes, phases, and derived quantities such as the frequency detuning $\delta \nu$ and the relative phase $\Phi$, next to the extracted frequencies. The strong frequency-phase covariances detected in all covariance matrices displayed in Appendix~A.4 therefore render it necessary to include such correlations and covariances in future nonlinear asteroseismic modelling of the SPB stars in our sample.

\begin{figure*}[ht]
    \centering
    \includegraphics[width=18cm]{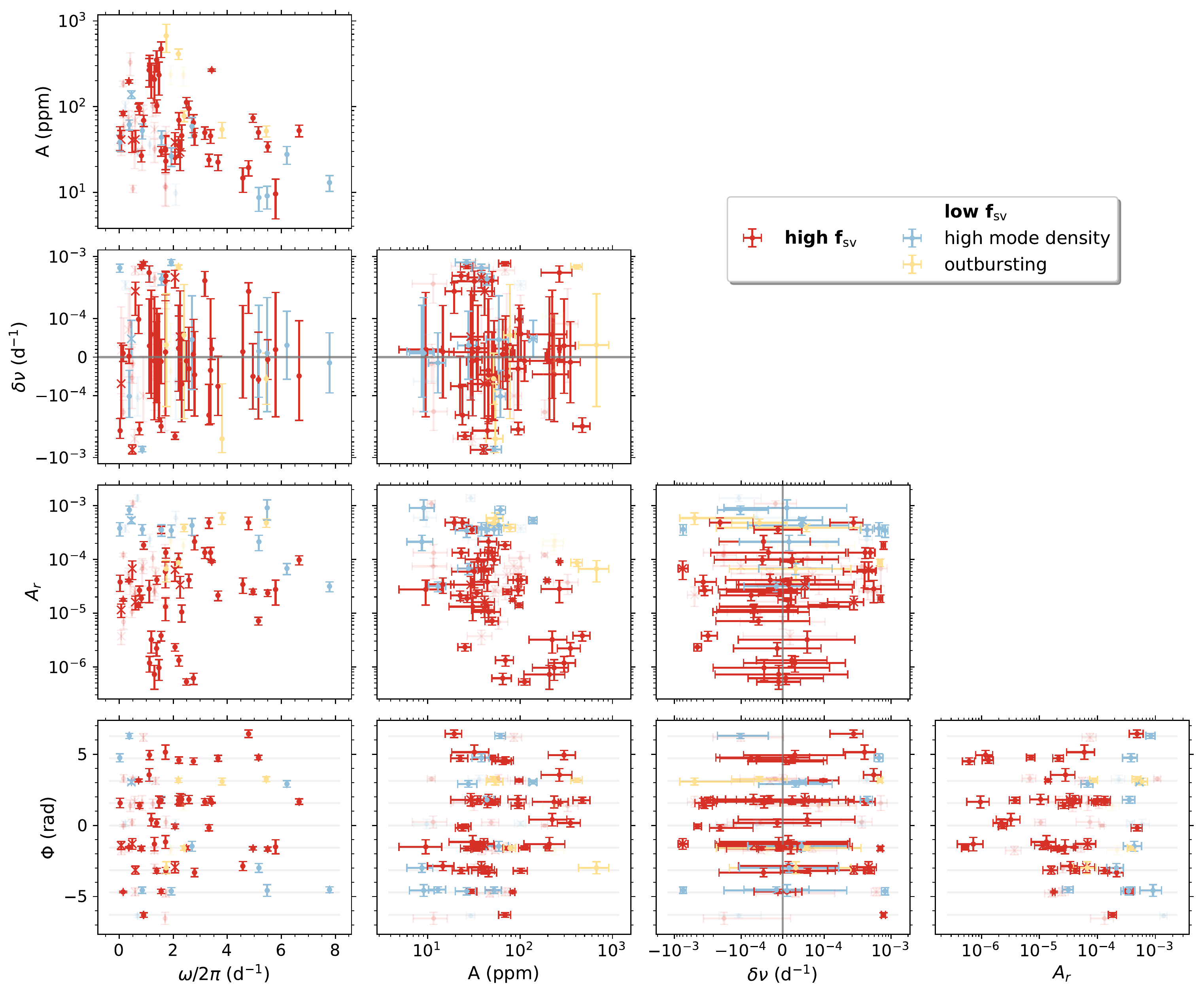}
    \caption{Highest-$f_{\rm sv}$ strategy characteristic properties of the identified robust candidate resonances, which involve the two largest-amplitude signals, as a function of each other, with $1$-$\sigma$ uncertainties indicated: amplitude $A$, frequency $\omega/2\pi$, frequency detuning $\delta\nu$ (see Eqn.~(\ref{eq:frequency_relation})), relative amplitude $A_r$ (see Eqn.~(\ref{eq:relative_amplitude})), and relative phase $\Phi$ (see Eqn.~(\ref{eq:relative_phase_relation})). Colours indicate the different pseudoclasses to which the stars that exhibit the candidate resonances belong. The transparent symbols indicate robust candidate resonances having $4<$ S/N $< 5$, whereas the full symbols indicate robust candidate resonances for which S/N $\geq 5$. Grey lines indicate the zero-point for the frequency detuning, and the faint grey lines in the last row of panels indicate the relative phases satisfying Eqn.~(\ref{eq:relative_phase_relation}).}
    \label{fig:nonlinear_characteristics_two_largest_amplitudes}
\end{figure*}

No differences in the number of identified two-signal candidate resonances are found when comparing the models of stars belonging to the low- and high-$f_{\rm sv}$ pseudoclasses, as is shown in the left panel of Fig.~\ref{fig:comparison_s_f_v_robust_resonance}. This conclusion is not changed when only considering two-signal candidate resonances with S/N $\geq 5$. The box-and-whisker plots in that panel furthermore show that no significant dependence on prewhitening strategy can be identified. Any additional uncertainty in parameter values due to covariance among parameters is neglected in the construction of the box-and-whisker plots. However, including such uncertainties in their construction would only support the conclusion that no significant dependencies are observed.

One might suspect that the differing (rich) objective function landscape for which the global minimum is sought during optimisation is a reason for a difference between stars belonging to the two pseudoclasses, high and low $f_{\rm sv}$. Amplitude modulation is more important in shaping this landscape for low-$f_{\rm sv}$ stars than their high-$f_{\rm sv}$ counterparts, although this varies from star to star and from signal to signal. However, no difference in number of identified two-signal robust candidate resonances (i.e.\ those that fulfil the condition in Eqn.~(\ref{eq:robust_match})) can be noted when comparing the two pseudoclasses, as is shown on the right panel of Fig.~\ref{fig:comparison_s_f_v_robust_resonance}. This lack of difference is also noted for robust two-signal candidate resonances with S/N $\geq 5$. Besides a few exceptions, amplitude modulation of low-$f_{\rm sv}$ star signals thus seems to have only a minor influence on the identification of signals relevant for (two-signal) robust candidate resonances.

We find no evidence for a distinction in number of identified candidate resonances when comparing the numbers strategy per strategy for outbursting and non-outbursting stars, even if we only compare candidate resonances for which S/N $\geq 5$. A class imbalance exists when making the latter comparison (8 outbursting stars vs. 30 non-outbursting stars).
However, the presence of robust candidate resonances in stars of all (pseudo)classes supports the lack of distinction between outbursting and non-outbursting stars. It also lends support to the interpretation of Be star variability in which the times of outbursts signify several modes coming into phase \citep[e.g.,][]{2015MNRAS.450.3015K}.

Finally, we also check for correlation of the number of candidate resonances with binarity. Two stars in our sample are confirmed binaries, KIC006352430 \citep{2013A&A...553A.127P} and KIC004930889 \citep{2017A&A...598A..74P}, whereas \citet{2011A&A...526A.124L} suspect KIC008459899 to be a binary as well. All other stars in our sample do not have confirmed or suspected binary companions. We find no difference in the number of identified two-signal candidate resonances between (suspected) binary and single SPB stars, and only including candidate resonances for which S/N $\geq 5$ does not make a difference. 

\subsection{Characteristic properties of identified candidate resonances}\label{subsubsec:characteristics_candidate_resonances}

Amplitudes, frequencies, frequency detunings $\delta\nu$, relative amplitudes $A_r$, and relative phases $\Phi$ define candidate resonances. These characteristic properties are for example discussed by
\citet{2009A&A...506..111D} for a $\beta$ Cep (CoRoT) star, HD 180642, whereas \citet{2000MNRAS.313..179V} and \citet{2000MNRAS.313..185V} discuss them for an oscillating DA white dwarf, G29-38. We present how these properties covary in Fig.~\ref{fig:nonlinear_characteristics_two_largest_amplitudes} for the identified robust two-signal candidate resonances of all stars in our sample. The amplitudes and frequencies of signals involved in these resonances are not significantly different from those not involved in them, as shown in Fig.~\ref{fig:distribution_plot_scarcity_strategies}.

Most robust (two-signal) candidate resonances originate from high $f_{\rm sv}$ stars, as can be seen in Table \ref{tab:2_largest_amplitude_frequencies_count_overview_strategies}. The panels in the left column of Fig.~\ref{fig:nonlinear_characteristics_two_largest_amplitudes} show that most robust two-signal candidate resonances exhibit frequencies $\lesssim 2.5$ d$^{-1}$. Several of those low-frequency ($\omega/2\pi \lesssim 2.5$ d$^{-1}$) candidate resonances have $4<$ S/N $< 5$. The bulk of the high-frequency robust two-signal candidate resonances ($\omega/2\pi \gtrsim 2.5$ d$^{-1}$) depicted in the top panel of Fig.~\ref{fig:nonlinear_characteristics_two_largest_amplitudes} exhibit amplitudes $< 100$ ppm, whereas the bulk of the low-frequency robust two-signal candidate resonances exhibit amplitudes varying from $\sim30$ to $\sim300$ ppm. Only a few of the detected robust two-signal candidate resonances have amplitudes $\geq 200$ ppm, predominantly at low frequencies. Similarly, only a few robust two-signal candidate resonances are detected with amplitudes of the order of $10$ ppm.
Few of the frequency detunings of the robust candidate resonances are significantly different from $0 $ d$^{-1}$, as can be deduced from the panels in Fig.~\ref{fig:nonlinear_characteristics_two_largest_amplitudes} that display the frequency detunings ($\delta\nu$). This is expected because of the way we select robust candidate resonances (see Eqn.~(\ref{eq:robust_match})). The few outliers we observe likely either belong to the frequency-locked or to the `intermediate' regime, although definitive conclusions can only be made if the linear growth rate $\gamma$ of the mode is known, which requires nonlinear asteroseismic models. No distinct trends are present in any of the panels of Fig.~\ref{fig:nonlinear_characteristics_two_largest_amplitudes} that display the relative phases ($\Phi$) and the relative amplitudes ($A_r$).

Variances for the characteristic properties are larger than average for robust two-signal candidate resonances that have extreme amplitudes.
By comparing the individual properties of the two-signal candidate resonances for each star in our sample, we find that the largest-amplitude signals shown in Fig.~\ref{fig:nonlinear_characteristics_two_largest_amplitudes} are the robust two-signal candidate resonances of KIC009715425 and KIC010526294. The larger uncertainties in their properties are attributable to the high mode density for the outbursting star KIC009715425, and the unresolved rotationally split triplet signals in the light curve of KIC010526294 \citep{2014A&A...570A...8P}. The covariance matrices of the best light curve regression models of these stars are strongly structured, as shown in Figs~A.6 and A.7, indicating significant covariance among some of the regression parameters. For the small-amplitude signals shown in Fig.~\ref{fig:nonlinear_characteristics_two_largest_amplitudes}, the increased variance of the characteristic properties is explained by the fact that the signals are closer to the noise level, as is indicated by half of these signals having $4<$ S/N $< 5$.

The approach taken in this section to characterise candidate resonances contains the conservative approximation of cataloguing only the candidate resonances influenced by the two largest-amplitude signals.
It is thus limited in capturing the characteristic properties of candidate resonances. This raises questions that need to be solved: 
\begin{enumerate}
    \itemsep0.3em
    \item  Is the conservative approach that only considers robust candidate resonances defined by Eqn.~(\ref{eq:robust_match}) satisfactory for initial target selection for nonlinear asteroseismic studies?
    \item Assuming the conservative approach is satisfactory, how do the number and characteristic properties of candidate resonances change if more, lower-amplitude signals are allowed to be involved in the candidate resonances?
    \item Is it expected from nonlinear stellar oscillation theory that no significant trends exist among the characteristic candidate resonance properties?
\end{enumerate}
The second question is dealt with in Appendix B, whereas the other questions require nonlinear asteroseismic models in order to be answered, which we plan to take up in follow-up studies.

\section{Conclusions and prospects}\label{sec:conclusions_prospects}

We analysed the {\em Kepler\/} light curves of an ensemble of $38$ SPB stars
by making use of five different iterative prewhitening strategies.
Amplitude-based prewhitening strategies tend to extract larger-amplitude signals from the lower-frequency, mode-dense regions of the LS periodograms in $g$-mode pulsators, whereas S/N-based prewhitening strategies tend to extract more smaller-amplitude signals at higher-frequency, mode-scarce regions, as illustrated in Fig.~\ref{fig:distribution_plot_scarcity_strategies}. 

The scaled fraction of variance $f_{\rm sv}$, defined in Eqn.~(\ref{eq:scaled_fraction_variance}), is one of the metrics used to compare the frequency analysis results, in order to select the optimal model for the light curve. Based on that metric, as well as the (residual) LS periodogram and light curve, we divide the stars in our sample in three different pseudoclasses: the high $f_{\rm sv}$, the high mode density, and the outbursting pseudoclass. The first pseudoclass consists of $19$ stars 
for which the models of all five prewhitening strategies attain $f_{\rm sv}$ values $> 0.9$. The two other pseudoclasses are discerned by detection of outburst-like features. We refer to them as pseudoclasses because we found that some of the characterising properties of these pseudoclasses are attributable to the limited frequency resolution that does not allow one to resolve several of the signals observed in the {\em Kepler} light curves. This limited frequency resolution can be attributed to the high mode densities that are detected in the LS periodograms, which also leads to modulation of some of the detected signals.
The resonant excitation of envelope $g$ modes with inertial modes propagating in the core, which requires rapid rotation to be efficient
\citep{2021MNRAS.505.1495L}, may also be a reason why SPB stars exhibit such mode-dense LS periodograms.

An important point in the frequency analysis of SPB stars is the covariance among the fitting parameters. We demonstrate the particularly strong covariances between extracted frequencies and phases in the figures in appendix A. These covariances are sometimes neglected in the uncertainty budget of the light curve models, so that regression parameter uncertainties are underestimated. It is necessary to include such regression parameter covariations in future asteroseismic modelling of the SPB stars in our sample.

The role of resonances among $g$ modes was described eloquently by \citet{1997A&A...321..159B}. Few to no space-based photometric data were available at the time to match the mainly theoretical discussions of its importance. \citet{2012MNRAS.420.2387L} provided account of the importance of rotation on nonlinear mode interactions in the rapidly rotating B stars by studying $g$ modes excited in stellar models near the zero-age main sequence.
Applications of this theory were limited by the lack of available photometric data. Because we now have the full {\em Kepler} data set at our disposal, the research field of nonlinear asteroseismology applied to main sequence $g$-mode pulsators can be reinvigorated. Although intermediate-mass dwarfs  have primarily been the target of linear asteroseismic studies for widely differing purposes, such as 
their sizing, weighing, ageing \citep{2021A&ARv..29....4S}, as well as the characterisation of their internal rotation and mixing \citep[e.g.,][]{2021NatAs.tmp...80P, 2021A&A...650A.175M}, no observation-based modelling studies have focused on nonlinear mode interactions.

Given the observed deficit of angular momentum transport during the main sequence in current stellar evolution models \citep{2019ARA&A..57...35A}, the importance of nonlinear mode interactions for angular momentum transport deserves follow-up studies. In this work we provide an observational basis for this.
Based on the light curve models constructed during frequency analysis, we identified second-order candidate direct resonant {\em Kepler} SPB oscillations. We guided ourselves in this identification process by the frequency detuning and relative phase relations defined in Eqns~(\ref{eq:frequency_relation}) and (\ref{eq:relative_phase_relation}), inspired by \citet{1995A&A...295..371V} and \citet{1997A&A...321..159B}. Future nonlinear asteroseismic models describing frequency-locked signals based on the amplitude equations, will make use of such characteristic properties of candidate resonances (as well as the relative amplitude defined in Eqn.~(\ref{eq:relative_amplitude})), or derivates thereof. 

It is more likely that a resonant $g$ mode interaction occurs if one of the parent frequencies has a large amplitude.
As a first approximation we look at the number of identified candidate resonances that involve the two largest-amplitude signals in the light curve, which we refer to as `two-signal' candidate resonances, and the characteristic properties of such resonances. We find that 32 of the 38 SPB stars turn out to be intrinsic nonlinear $g$-mode oscillators, reiterating the need for quantification of the nonlinear $g$ mode interactions. 

One should take into account that the practical implementation of the criterion indicating that the frequency detuning relation is fulfilled is only approximate, because the growth rates needed to verify the \citet{1997A&A...321..159B} criterion cannot be derived directly from observations. Moreover, as \citet{1998BaltA...7...21G} explain, the boundaries of the \citet{1997A&A...321..159B} regimes of resonant mode interaction might not be that distinct for observational data, and they are undetermined for SPB stars A step towards the determination of the  \citet{1997A&A...321..159B} regimes consists of verifying whether the candidate resonances detected in this work display the oscillatory behaviour that is deemed characteristic of the `intermediate regime' \citep{1998BaltA...7...21G}. This is a next logical step in characterising $g$ mode resonances in SPB stars. 

Our current observational analyses constitute the ground work to perform forward asteroseismic modelling of the nonlinear mode properties, with the aim to investigate whether it leads to different or additional results on internal rotation and mixing compared to the current state-of-the-art based on the linear asteroseismic approach in stellar modelling of SPB stars \citep{2021NatAs.tmp...80P}. 

\begin{acknowledgements} The research leading to these results received funding from
the European Research Council (ERC) under the European Union’s Horizon
2020 research and innovation program (grant agreement No. 670519: MAMSIE with PI CA.), and from the KU Leuven Research Council (grant C16/18/005: PARADISE, with PI CA.). DMB is grateful for a senior post-doctoral fellowship from the Research Foundation Flanders (FWO) with grant agreement No. 1286521N. TVR gratefully acknowledges support from the Research Foundation Flanders (FWO) under grant agreement nr. 12ZB620N. This research was supported in part by the National Science Foundation under Grant No. NSF PHY-1748958. A part of the computational resources and services used in this work were provided by the VSC (Flemish Supercomputer Center), funded by the Research Foundation - Flanders (FWO) and the Flemish Government – department EWI. JVB thanks J. De Ridder for his advice on visualising correlations among light curve regression model parameters, and thanks W. Zong for his useful comments on the manuscript. Throughout this work we have made use of the following Python packages: {\sc lmfit} \citep{matt_newville_2020_3814709}, {\sc SciPy} \citep{2020SciPy-NMeth}, {\sc NumPy} \citep{2020NumPy-Array}, Pandas \citep{mckinney-proc-scipy-2010}, {\sc Matplotlib} \citep{4160265}, and {\sc Seaborn} \citep{waskom2020seaborn}. We thank their authors for making these great software packages open source. \end{acknowledgements}

\bibliographystyle{aa.bst} 
\bibliography{ourbiblio.bib} 

\end{document}